\documentclass[aps,prd,preprintnumbers,groupedaddress,nofootinbib,amssymb,notitlepage,eqsecnum]{revtex4-2}
\usepackage{here}
\usepackage[dvipdfmx]{graphicx}
\usepackage{amsmath,amsthm,amssymb}
\usepackage{bm}
\usepackage{color}
\usepackage[dvipsnames]{xcolor}
\usepackage[normalem]{ulem}

\usepackage{amsfonts}
\usepackage{dcolumn}
\usepackage[colorlinks=true,hyperfootnotes=true,citecolor=cyan]{hyperref}
\allowdisplaybreaks[1]
\usepackage{stackengine}

\newcommand{\be}{\begin{equation}}  
\newcommand{\ee}{\end{equation}}
\newcommand{\ba}{\begin{eqnarray}}
\newcommand{\ea}{\end{eqnarray}}

\newcommand{\rd}{{\rm d}}

\newcommand{\bem}{\begin{bmatrix}}
\newcommand{\eem}{\end{bmatrix}}

\newcommand{\mK}{\mathcal{K}}

\newcommand{\mS}{\mathcal{S}}

\newcommand{\Lag}{\mathcal{L}}
\newcommand{\dd}{\text{d}}

\newcommand{\mKCh}{\mathcal{K}_{\text{gChs}}}
\newcommand{\rhoD}{\rho_{\text{D}}}
\newcommand{\PD}{P_{\text{D}}}
\newcommand{\kL}{k_{\text{L}}}

\allowdisplaybreaks

\begin{document}

\preprint{WUCG-26-05}

\title{A solid unification of the dark sector}

\author{Jose Beltr\'an Jim\'enez$^{1,2}$\footnote{{\tt 
jose.beltran@usal.es}}}
\author{Mar\'ia P\'erez Garrote$^{1,2}$\footnote{{\tt 
maripe49@usal.es}}}
\author{Florencia A. Teppa Pannia$^{2,3}$\footnote{{\tt 
fa.teppa.pannia@upm.es}}}
\author{Shinji Tsujikawa$^{4}$\footnote{{\tt 
tsujikawa@waseda.jp}}}

\affiliation{$^1$Departamento de F\'isica Fundamental, Universidad de Salamanca, E-37008 Salamanca, Spain.\\
$^2$Instituto Universitario de F\'isica Fundamental y Matem\'aticas IUFFyM, Universidad de Salamanca, E-37008 Salamanca, Spain.\\
$^3$Departamento de Matem{\'a}tica Aplicada a la Ingenier{\'i}a Industrial, Universidad Polit{\'e}cnica de Madrid, E-28006 Madrid, Spain.\\
$^4$Department of Physics, Waseda University, 
3-4-1 Okubo, Shinjuku, Tokyo 169-8555, Japan.}

\begin{abstract}
We construct a unified description of dark matter and dark energy in terms of a single dark component that behaves as a pressureless fluid in the early Universe and undergoes a transition to a solid phase at late times that can support accelerated expansion. A generalized Chaplygin-type solid provides a simple realization of this scenario. The solid nature of the dark medium prevents the instabilities and strong acoustic oscillations that typically arise in perfect-fluid unifications. It also gives rise to distinctive signatures in cosmological perturbations, such as a suppression of structure growth, a nontrivial gravitational slip, and an effective mass for gravitational waves. Since these effects share a common origin in the solid phase, they become relevant only at low redshifts, leaving the high-redshift cosmology essentially unmodified. This demonstrates that a solidly unified dark sector can reproduce the desired background transition from dark matter to dark energy while yielding testable imprints in cosmological perturbations.
\end{abstract}

\date{\today}

%\pacs{04.50.Kd, 95.36.+x, 98.80.-k}

\maketitle

%%%%%%%%%%%%%%%%%%%%%%%%%%%%%%%%%
\section{Introduction}
\label{introsec}
%%%%%%%%%%%%%%%%%%%%%%%%%%%%%%%%%

The existence of dark matter (DM) and dark energy (DE) is one of the
most robust pieces of evidence that the standard model of particle
physics and general relativity (GR), in their simplest forms, do not yet
provide a complete description of the Universe.  Nonluminous matter was
first inferred from the dynamics of galaxies and clusters
\cite{Zwicky:1933gu,Rubin:1970zza}, and its gravitational role has since
been established by a broad set of observations, including cosmic
microwave background (CMB) anisotropies, large-scale structure, 
and gravitational lensing by galaxy clusters
\cite{WMAP:2012nax,Planck:2018vyg,Eisenstein:2005su,
Percival:2006gt,Tyson:1998vp,Clowe:2006eq,Bertone:2004pz}.
The accelerated expansion of the late Universe, discovered through Type
Ia supernovae \cite{Riess:1998cb,Perlmutter:1998np}, requires either a
new component with negative pressure or a modification of gravity on
cosmological scales.

The $\Lambda$-Cold-Dark-Matter ($\Lambda$CDM) model provides the minimal phenomenological framework for describing these 
observations, but it leaves unanswered 
the microscopic origin of both the cosmological constant and the dark matter 
component \cite{Weinberg:1988cp,Sahni:1999gb,Carroll:2000fy,Padmanabhan:2002ji,Peebles:2002gy,Martin:2012bt}.
Moreover, the persistence of the $H_0$ tension
\cite{Planck:2018vyg,Verde:2019ivm,DiValentino:2021izs,
Riess:2021jrx,Freedman:2021ahq} and recent DESI baryon acoustic oscillation (BAO) data, which show hints of deviations from a cosmological
constant when combined with CMB and supernova measurements
\cite{DESI:2024mwx,DESI:2025zgx}, motivate the exploration of dark-sector physics
beyond the minimal scenario.  
Dynamical DE models and 
modified gravity theories have therefore been widely investigated 
as possible extensions of $\Lambda$CDM
\cite{Fujii:1982ms,Peebles:1987ek,Wetterich:1987fm,Chiba:1997ej,Copeland:1997et,Caldwell:1997ii,Copeland:2006wr,Silvestri:2009hh,DeFelice:2010aj,Clifton:2011jh,Tsujikawa:2013fta,Joyce:2014kja,Koyama:2015vza,Kase:2018aps}.
Future surveys such as Euclid and the Legacy Survey of Space and Time
at the Vera Rubin Observatory will measure the cosmic expansion and
growth histories with increasing precision \cite{Euclid:2019clj,Euclid:2024yrr,LSSTDarkEnergyScience:2012szw,LSSTDarkEnergyScience:2018jkl}.
To exploit this information, it is useful to construct theoretically
controlled models in which the background evolution and perturbative
signatures of the dark sector can be predicted systematically.

In the conventional approach, DM and DE are treated as two separate
components: a pressureless clustering fluid and a nearly homogeneous negative-pressure component.  This separation, however, is not logically
mandatory.  A long-standing alternative is to unify DM and DE into a
single dark fluid whose equation of state interpolates between
matter-like behavior at early times and DE-like behavior at late times.
The Chaplygin gas and its generalized extensions provide prototypical
realizations of this idea
\cite{Kamenshchik:2001cp,Bilic:2001cg,Bento:2002ps,Dev:2002qa,Fabris:2001tc,Gorini:2002kf}.
At the background level, such models can reproduce a matter-dominated
era followed by cosmic acceleration.  The problem arises at the level of perturbations.  
In perfect-fluid generalized Chaplygin gas 
models, the adiabatic sound speed typically grows when the component starts to drive cosmic 
acceleration. The resulting pressure support either produces oscillatory
features or an excessive suppression or enhancement of the matter power spectrum.  This is why observations constrain the generalized Chaplygin gas very close to its $\Lambda$CDM limit \cite{Sandvik:2002jz}.  This difficulty is not a failure of the unification idea itself, but rather a consequence of
assuming that the unified medium is a perfect fluid with a single
adiabatic sound speed.

A natural way to go beyond the perfect-fluid description is to endow the
medium with rigidity.  Relativistic solids can support shear
deformations in addition to compression, and hence possess perturbative
responses that are absent in perfect fluids.  The effective field theory
(EFT) of relativistic media can be formulated in terms of three scalar fields that label the comoving volume elements of the medium
\cite{Dubovsky:2005xd,Endlich:2010hf,Dubovsky:2011sj,Endlich:2012pz,
Ballesteros:2012dm,Ballesteros:2014sxa,Kovtun:2014hpa}. The internal shift and rotational symmetries guarantee homogeneity and isotropy, while the state of the medium can be characterized by the overall compression and the two independent shear distortions at fixed volume. A perfect fluid is recovered when the theory has the larger symmetry of internal volume-preserving diffeomorphisms, in which case the Lagrangian depends only on the volume element so only the compressional mode remains.

This EFT language was developed in the study of relativistic fluids and
solids and was later used to build solid inflation
\cite{Endlich:2012pz}.  Subsequent cosmological applications include
statistical anisotropy and primordial non-Gaussianities in solid or
supersolid inflation
\cite{Bartolo:2014hwa,Ricciardone:2016lym,Celoria:2020diz,Celoria:2021cxq}.
It has also been applied to effective perfect fluids and higher-derivative
fluid corrections in cosmology \cite{Ballesteros:2012dm,Ballesteros:2014sxa},
and to the EFT of coupled DE and DM, where pressureless DM is treated as
a particular gravitating continuum \cite{Aoki:2025dmdeeft}.  More recent
formulations of gravitating media have clarified the relation among
solids, fluids, and other media from a unified symmetry perspective
\cite{Aoki:2022gez}.  These developments suggest that replacing a perfect
unified dark fluid by a solid medium may alter the perturbation dynamics
in precisely the regime where the Chaplygin gas encounters its strongest
constraints.

Several works have explored the possibility that DE itself may behave as
an elastic or solid component.  Relativistic elastic DE models and their
cosmological perturbations were studied in
Refs.~\cite{Battye:2005hw,Battye:2007aa}, and anisotropic solid DE
realizations have also been analyzed \cite{MotoaManzano:2020mwe}.  In
most of these studies, however, DM is still introduced as an independent
component, while the solid property is assigned only to the DE sector.
Thus, they are not unified DE-DM models, nor do they describe the entire
dark sector within a single EFT of a relativistic medium based on compression and shear 
invariants. The goal of the present work is to
fill this gap by constructing a single solid dark sector that behaves as
DM in the early Universe and as DE at late times.

We propose a concrete realization of such a solidly unified dark sector. The model exhibits a generalized Chaplygin-type background evolution. However, its perturbations are controlled not only by the compression mode but also by the shear response of the medium. This distinction is crucial. The background can reproduce the successful interpolation achieved by unified Chaplygin-type fluids, while the perturbations need not inherit the large perfect-fluid sound speed that would otherwise erase structure. Instead, the vector and longitudinal sound speeds are independent quantities determined by the solid response. Viable parameter regions can then be selected in which pressure support remains sufficiently small on the scales relevant to structure formation. In this sense, the model reformulates the usual Chaplygin-gas difficulty as a question about the elastic response of the dark medium, rather than as an unavoidable obstruction to dark-sector unification.

The framework developed here has several distinctive theoretical and
observational features.  First, we derive the stability conditions in the
tensor, vector, and scalar sectors. Tensor modes remain luminal but
acquire a time-dependent mass due to the shear rigidity of the solid, while vector and longitudinal scalar
phonons propagate with speeds fixed by the solid response.  Second, the
same solid structure that modifies the scalar sector also generates a
nonzero gravitational slip and an effective mass for gravitational waves (GWs),
providing signatures absent in perfect-fluid Chaplygin models.  Third,
we identify the parameter region in which the longitudinal pressure term
and the weakening of the effective gravitational attraction do not
excessively suppress the growth of dark-sector perturbations.  These
features make the model a useful laboratory for testing whether the dark
sector can be unified without reducing to a perfect fluid.

This paper is organized as follows. In Sec.~\ref{backsec}, we introduce the relativistic solid action and study the background dynamics of a generalized Chaplygin-type solid that interpolates between a matter era and cosmic acceleration. In Sec.~\ref{sec:stability}, we derive the linear stability conditions for tensor, vector, and scalar perturbations. In Sec.~\ref{perturbsec}, we obtain the linear scalar perturbation equations in a gauge-ready form and discuss the growth of the longitudinal perturbation and the associated dark-sector density contrast, clarifying the role of the vector and longitudinal sound speeds. In Sec.~\ref{signaturemodel}, we study observational signatures of the model, including  the longitudinal transfer function, the gravitational slip, 
and the GW distance. 
Section~\ref{conclusionsec} is devoted to conclusions.

%%%%%%%%%%%%%%%%%%%%%%%%%%%%%%%%%%%%%%%%%%%%
\section{The unified solid dark sector}
\label{backsec}
%%%%%%%%%%%%%%%%%%%%%%%%%%%%%%%%%%%%%%%%%%%%%%

\subsection{Theoretical framework}

We consider a solid, with a fluid included as a special case, to provide a unified description of DM and DE. The state of the solid is specified by tracking the motion of its individual volume elements. In a coordinate system $x^\mu=(t,\vec{x})$, 
we assign a three-dimensional comoving label $\phi^a$ (with $a=1,2,3$) to each volume element and describe its trajectory in physical space as $\vec{x}(\phi^a,t)$ \cite{Dubovsky:2005xd,Endlich:2012pz}. Equivalently,
we can use the inverse map and regard the labels $\phi^a(\vec{x},t)$ as scalar fields defined on spacetime.

The internal homogeneity of the medium is implemented by requiring that the underlying theory be invariant under a shift symmetry $\phi^a \to \phi^a+c^a$, with $c^a$ a constant internal vector. At lowest order in derivatives, the basic Lorentz-invariant building block is the internal matrix
\be
B^{ab} \equiv g^{\mu\nu}\partial_\mu \phi^a \partial_\nu \phi^b\,,
\ee
where $g^{\mu\nu}$ is the inverse spacetime metric tensor. Imposing invariance under internal SO(3) rotations, the Lagrangian can depend on three independent scalars constructed from $B^{ab}$ that can be conveniently chosen as \cite{Endlich:2012pz}
\be
X \equiv [B],\qquad
Y \equiv \frac{[B^2]}{[B]^2},\qquad
Z \equiv \frac{[B^3]}{[B]^3},
\label{XYZ}
\ee
where $[\cdots]$ denotes the trace over internal indices. The 
dark-sector Lagrangian can therefore be written as a function of
$X$, $Y$, and $Z$.

We are interested in a unified DM-DE scenario in which the medium behaves
as fluid-like DM in the early Universe and as solid-like DE at late times.
A defining property of a fluid is that volume elements can be internally
rearranged without changing the energy, as long as the total volume is
preserved. In the present language, this property corresponds to the
invariance of the dynamics under internal volume-preserving
diffeomorphisms,
$\phi^a \to \xi^a(\phi)$ with
$\det(\partial \xi^a/\partial \phi^b)=1$.
Among the three SO(3) invariants given in Eq.~(\ref{XYZ}), there is a
particular combination consistent with volume-preserving diffeomorphisms,
namely the determinant of $B^{ab}$:
\be
b \equiv \det B^{ab}
= \frac{X^3}{6}\left(1-3Y+2Z\right).
\ee
To describe our unified DM-DE scenario, it is thus convenient to express the dark-sector Lagrangian as a function of $b$, $Y$, and $Z$:
\be
{\cal L}_{\rm D}=-{\cal K}(b,Y,Z).
\ee
The invariant $b$ measures the local volume change of the medium and is therefore associated with the density, or compression, degree of freedom. By contrast, the dimensionless invariants $Y$ and $Z$ are insensitive to an overall rescaling of $B^{ab}$ and encode its shape after the volume change has been factored out. These invariants therefore characterize deformations at fixed volume, corresponding to shear distortions of a volume element.
In the fluid limit, where the theory is invariant under volume-preserving diffeomorphisms, the Lagrangian ${\cal L}_{\rm D}$ depends only on $b$ and deviations from the fluid regime are characterized by the dependence of ${\cal L}_{\rm D}$ on $Y$ and $Z$.

For the gravitational sector we will assume the standard Einstein-Hilbert Lagrangian $R/(16\pi G)$ of GR, where $R$ is the Ricci scalar and 
$G$ is Newton's constant. 
To describe matter components other than the dark sector, such as baryons and radiation, we include a matter Lagrangian $\Lag_{\text{M}}$. 
The total action can then be written
\be
{\cal S}=\int {\rm d}^4 x \sqrt{-g}
\left[ \frac{R}{16 \pi G}-{\cal K}(b,Y,Z)+\Lag_{\text{M}}\right]\,,
\label{action}
\ee
where $g$ is the determinant of 
the metric tensor $g_{\mu \nu}$.

The idea behind the construction of the unified dark sector is to choose a function $\mK$ whose dynamics are dominated by the dependence on $b$ at high redshift, while the effects associated with $Y$ and $Z$ become relevant only at late times. In other words, we require a hierarchy in which the contributions from the $Y$ and $Z$ dependence are subdominant compared with those from the $b$ dependence at high redshift, but this hierarchy is gradually relaxed as the Universe expands. In this regime, the medium behaves to a good approximation as a perfect fluid at early times, whereas solid effects become important at late times. The model thus describes a transition from a fluid phase to a solid phase. The question is whether this transition can arise dynamically in a natural way, or whether it requires a contrived construction. In the next subsection, we show that the desired behavior is realized.

\subsection{Background equations of motion}

To show that a transition from a fluid phase to a solid phase can be achieved in the cosmological evolution, let us study the cosmological dynamics on a spatially flat Friedmann-Lema\^itre-Robertson-Walker (FLRW) background, whose line element is given by
\be
\mathrm{d}s^2 = -\mathrm{d}t^2 + a^2(t)\,\delta_{ij}\,
\mathrm{d}x^i \mathrm{d}x^j,
\label{backmet}
\ee
with $a$ the scale factor. The appropriate configuration for $\phi^a$ to support the homogeneous and isotropic FLRW metric is the following
\be
\bar{\phi}^a=\lambda x^a\,,
\label{phiaback}
\ee
where $\lambda$ is a constant parameter. 
On the background configuration (\ref{backmet}), 
we have
\be
b = \frac{\lambda^6}{a^6} \,, \qquad 
Y = \frac{1}{3} \,, \qquad 
Z = \frac{1}{9} \,,
\label{backXYZ}
\ee
so that only $b$ is time dependent.
The constant parameter $\lambda$ can be absorbed into a rescaling of the spatial coordinates.
Therefore, without loss of generality, 
we can set $\lambda=1$, for which
$b=1/a^6$. 

Although the above configuration is neither isotropic nor homogeneous, it is compatible with the spatially flat FLRW background, as it respects its homogeneity and isotropy thanks to the assumed internal symmetries. Indeed, the internal shifts can be used to compensate a spatial translation, while the internal SO(3) symmetry permits to compensate a spatial rotation. We can verify that the above configuration is compatible with homogeneity and isotropy by explicitly computing the energy-momentum tensor, which sources the Einstein equations. 
The energy-momentum tensor is given by
\be
T_{\text{D},\mu\nu}
\equiv -\frac{2}{\sqrt{-g}}\frac{\delta \mS_{\text{D}}}{\delta g^{\mu\nu}}
=2\frac{\partial \mK}{\partial B^{ab}}\partial_\mu\phi^a\partial_\nu\phi^b-g_{\mu\nu}\mK.
\label{T solid}
\ee
Using $\mK=\mK(b,Y,Z)$, this can be written as
\be
T_{\text{D},\mu\nu}
=2b\mK_{,b} B^{-1}_{ab}\partial_\mu\phi^a\partial_\nu\phi^b
+4\mK_{,Y}\left(\frac{B_{ab}}{X^2}-\frac{Y}{X}\delta_{ab}\right)\partial_\mu\phi^a\partial_\nu\phi^b
+6\mK_{,Z}\left(\frac{B^c{}_aB_{bc}}{X^3}-\frac{Z}{X}\delta_{ab}\right)\partial_\mu\phi^a\partial_\nu\phi^b-g_{\mu\nu}\mK\,,
\label{TDsolid}
\ee
where $B^{-1}_{ab}$ is the inverse of $B^{ab}$.
Since we have $B^{ab}=g^{ij}=a^{-2}\delta^{ij}$ on the FLRW background, we can easily understand why the above energy momentum does not give rise to any anisotropic contribution and is entirely described in terms of an energy density and isotropic pressure. 

We define the four-velocity of the medium 
$u^\mu$ through the conditions
$u^\mu\partial_\mu \phi^a=0$ 
and $u^\mu u^\nu g_{\mu\nu}=-1$.
The future-directed solution 
is \cite{Dubovsky:2005xd, Endlich:2010hf, Endlich:2012pz}
\be
u^\mu=
\frac{1}{6\sqrt{b}\sqrt{-g}}
\epsilon^{\mu\alpha\beta\gamma}
\epsilon_{abc}
\partial_\alpha\phi^a
\partial_\beta\phi^b
\partial_\gamma\phi^c\,,
\label{u_def}
\ee
where $\epsilon^{\mu\alpha\beta\gamma}$ is the four-dimensional Levi-Civita symbol, with $\epsilon^{0123}=+1$. The projector on hypersurfaces orthogonal to 
the four-velocity is
\be
h_{\mu\nu}=g_{\mu\nu}+u_\mu u_\nu,\qquad h_{\mu\nu}u^\nu=0.
\label{h_def}
\ee
The energy density and isotropic pressure of the medium are obtained from
\be
\rho_{\rm D}=T_{{\rm D},\mu\nu}u^\mu u^\nu,\qquad
P_{\rm D}=\frac{1}{3}h^{\mu\nu}T_{{\rm D},\mu\nu}.
\ee
Using Eq.~(\ref{TDsolid}), this gives
\be
\rho_{\rm D}={\cal K},\qquad
P_{\rm D}=2b{\cal K}_{,b}-{\cal K}\,.
\label{rhoPsolid}
\ee
From these relations, we find that 
$\rho_{\rm D}+P_{\rm D}=2b\mK_{,b}$, 
so the null energy condition is satisfied 
provided $2b\mK_{,b}>0$. 
We will see later that the same condition is required to avoid ghost instabilities in our scenario. The null energy condition must therefore be satisfied by the model. 

The background energy density and pressure 
in the dark sector satisfy the usual continuity equation,
\be
\dot{\rho}_{\rm D}+3H \left( \rho_{\rm D}+P_{\rm D} \right)=0\,,
\label{continuityD}
\ee
where $H=\dot{a}/a$ is the Hubble expansion rate, and a dot denotes differentiation with respect to $t$. 
It is important to note that the background evolution depends only on derivatives of
${\cal K}$ with respect to $b$, and not on derivatives with respect to $Y$ or $Z$.
Thus, only $b$ is sensitive to the cosmological expansion, whereas $Y$ and $Z$ remain constant; see Eq.~\eqref{backXYZ}. This reflects the fact that $Y$ and $Z$ are defined in such a way that the overall expansion is factored out \cite{Endlich:2012pz}. As a result, the background evolution is governed only by the compression mode, even though the medium is a solid.

The equation-of-state parameter in the 
dark sector is
\be
w_{\rm D}=\frac{P_{\rm D}}{\rho_{\rm D}}=-1+\frac{2b{\cal K}_{,b}}{{\cal K}}\,.
\label{wD}
\ee
From this expression, accelerated quasi-de Sitter expansion can be supported when $2b{\cal K}_{,b}/{\cal K}\ll 1$. On the other hand, a dust-like evolution with $w_{\rm D}\simeq 0$ is obtained for ${\cal K}=m\sqrt{b}$, where $m$ is a constant. We can therefore construct a unified scenario by choosing a function ${\cal K}$ that interpolates between ${\cal K}=m\sqrt{b}$ at large $b$ and a form satisfying $2b{\cal K}_{,b}/{\cal K}\ll 1$ at small $b$. In the latter regime, a dependence on $Y$ and $Z$ is introduced to realize the transition to the solid phase. Since the cosmological evolution drives $b$ toward small values in an expanding Universe, while $Y$ and $Z$ remain constant, the transition from the pressureless fluid phase to the accelerating solid phase can occur dynamically, without relying on tuned properties of the background evolution.

For completeness, let us write the gravitational equations of motion. The Friedmann equations governing the background evolution are
\ba
& &
3H^2=8 \pi G \left( \rho_{\rm D}+\rho_{\rm M}\right)\,,
\label{back1} \\
& &
2\dot{H}+3H^2=-8 \pi G \big( P_{\rm D}+P_{\rm M}\big)\,,
\label{back2} 
\ea
where the matter sector has energy density $\rho_{\rm M}$ and pressure $P_{\rm M}$, and satisfies the usual continuity equation
\be
\dot{\rho}_{\rm M}+3H \left( \rho_{\rm M}+P_{\rm M} \right)=0\,.
\label{back0}
\ee
This matter contribution consists of all species other than the dark sector, such as radiation and baryons. 
The background energy densities of radiation and baryons evolve according to the usual scaling laws,
\be
\rho_{\text{r}}=\rho_0\Omega_{\text{r}} a^{-4},\qquad \rho_{\text{b}}=\rho_0\Omega_{\text{b}} a^{-3}\,,
\ee
where $\rho_0=3H_0^2/(8\pi G)$ is the critical density today, and $\Omega_{\text{r}}$ and $\Omega_{\text{b}}$ are the corresponding density parameters evaluated today.
We can similarly introduce the present-day density parameter of the dark sector, $\Omega_{\text{D}}\equiv\rho_{\text{D},0}/\rho_0$, so that the usual cosmic sum rule is 
\be
\Omega_{\text{D}}+\Omega_{\text{b}}+\Omega_{\text{r}}=1.
\ee

Combining the gravitational equations gives
\be
\dot{H}=-4\pi G \left( 2b {\cal K}_{,b}
+\rho_M+P_M \right)\,.
\label{back3} 
\ee
Taking the time derivative of this relation, 
we obtain
\be
\frac{\ddot{H}}{H}=12 \pi G H \Big[ 4b \left({\cal K}_{,b}+b{\cal K}_{,bb} \right)
+\left( c_{\rm M}^2+1 \right) \left( \rho_{\rm M}+P_{\rm M} \right) \Big]\,,
\ee
where we have introduced the adiabatic
sound speed squared of the matter sector,
\be
c_{\rm M}^2=\frac{\dot{P}_{\rm M}}{\dot{\rho}_{\rm M}}\,.
\label{cM}
\ee
We also define the effective equation-of-state parameter as
\be
w_{\rm eff}=-1+\frac{2}{3}\epsilon\,,
\label{weff}
\ee
where
\be
\epsilon \equiv -\frac{\dot{H}}{H^2}
=\frac{3(2b {\cal K}_{,b}+\rho_{\rm M}+P_{\rm M})}
{2({\cal K}+\rho_{\rm M})}\,.
\label{epsilon}
\ee
In deriving this expression, we have used Eqs.~(\ref{back1}) and (\ref{back3}). Since $\ddot{a}/a=H^2(1-\epsilon)$, the condition for cosmic acceleration, $\ddot{a}>0$, is equivalent to $\epsilon<1$, or $w_{\rm eff}<-1/3$. When the contributions from baryons and radiation are negligible compared to that of the solid dark sector in Eq.~(\ref{epsilon}), we have $\epsilon \simeq 3b{\cal K}_{,b}/{\cal K}$. This expression shows that a constant slow-roll evolution, with constant $\epsilon$, can be achieved for
\be
{\cal K} \simeq {\cal K}_0 b^{\epsilon/3}\,,
\label{Ksim}
\ee
where ${\cal K}_0$ is an integration constant. In the solid dark sector, the parameters ${\cal K}_0$ and $\epsilon$ can be promoted to functions of $Y$ and $Z$. Such a dependence does not affect the background evolution, since $Y$ and $Z$ are constant on the background, but it is crucial for the evolution of perturbations. This is the type of evolution we use for the accelerating solid phase of our unified dark sector. It should be kept in mind, however, that a much larger class of models is possible, since accelerated expansion only requires $\epsilon \ll 1$.

Having explained how a solid medium can accommodate a unified dark sector, it is instructive to present an explicit realization. It is important to keep in mind, however, that the specific model considered below is only a representative example, while the solid unification mechanism itself is much more general.

\subsection{Generalized Chaplygin solid}
\label{sec:GChs}

We consider an explicit realization of solid unification motivated by earlier attempts to unify the dark sector with a generalized Chaplygin gas \cite{Bento:2002ps}. Such scenarios were shown in Ref.~\cite{Sandvik:2002jz} to suffer from strong acoustic oscillations or instabilities. An interesting outcome of our construction is that these unwanted features can be tamed by the solid effects. The generalized Chaplygin solid considered here is described by
\be
\mKCh=m \sqrt{b} \left[ 
1+\left( \frac{b^{\epsilon_0/3}}
{m \sqrt{b}} \right)^{1+\alpha} 
f(Y,Z) \right]^{1/(1+\alpha)}\,,
\label{calK}
\ee
where $m$, $\alpha$, and $\epsilon_0$ are constants, and $f$ is a function of $Y$ and $Z$. We assume that $m$, $\epsilon_0$, and $f$ are positive, with $0<\epsilon_0\ll 1$. For $\alpha=0$, the function ${\cal K}$ can be cleanly decomposed into a dust-like DM contribution and a DE contribution. It is important to note, however, that these two contributions share the same degrees of freedom and therefore still constitute a unified dark sector, rather than two separate DM and DE components. For $\alpha\neq0$, such a decomposition cannot be performed, and the genuinely unified nature of the dark sector becomes more apparent. Since $Y$ and $Z$ are constant on the background, $f$ and its derivatives reduce to constant parameters.

\begin{figure}
\centering
\includegraphics[width=0.49\linewidth]{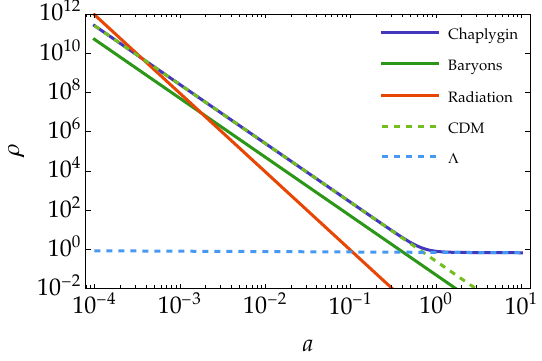}
\includegraphics[width=0.49\linewidth]{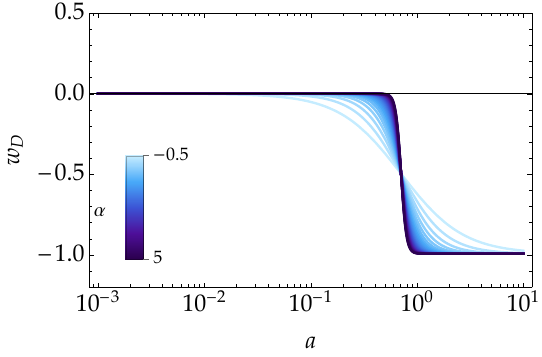}
\caption{In the left panel, we show the evolution of the energy densities as a function of the scale factor $a$ in the generalized Chaplygin solid model. 
The scale factor is normalized such that $a=1$ today.
We have chosen $\epsilon_0=0.01$, $\alpha=0.5$,
$\Omega_{\text{DM}}=0.25$, $\Omega_{\text{b}}=0.05$, $\Omega_{\text{r}}=9\times 10^{-5}$, and $\Omega_{\Lambda}=1-\Omega_{\text{DM}}-\Omega_{\text{b}}-\Omega_{\text{r}}$. The corresponding $\Lambda$CDM model with the same values of the cosmological parameters is shown by dashed lines. We see that the Chaplygin solid smoothly interpolates between the CDM and $\Lambda$ regimes. The main effect of $\alpha$ is to control the smoothness of this transition, as illustrated in the right panel, where we show the dark-sector equation of state for several values of $\alpha$.
}
\label{Fig:Chpdensities}
\end{figure}

It is instructive to rewrite Eq.~\eqref{calK} 
in terms of the cosmological parameters of 
the interpolating $\Lambda$CDM limit, which is recovered for
$\epsilon_0=\alpha=0$ and constant $f$. In this case, we have
$\mKCh=m\sqrt{b}+f$, so the energy density of the dark 
solid in Eq.~\eqref{rhoPsolid} on the cosmological background reduces to $\rhoD=\mKCh
=ma^{-3}+f(1/3,1/9)$, where we have set $\lambda=1$. 
Comparing this expression with the dark-sector
energy density in the standard $\Lambda$CDM model,
\be
\rho_{\Lambda{\rm CDM}}
=\rho_0\left(\Omega_{\rm DM}a^{-3}+\Omega_\Lambda\right)\,,
\ee
where $\Omega_{\rm DM}$ and
$\Omega_\Lambda$ are the present-day density 
parameters of dark matter and
the cosmological constant, respectively, it is natural to identify
$m=\rho_0\Omega_{\rm DM}$ and $f(1/3,1/9)=\rho_0\Omega_\Lambda$ in the
$\Lambda$CDM limit. For general $\alpha$, we instead normalize $f(1/3,1/9)$ as 
\be
f(1/3,1/9)=\left(\rho_0\Omega_\Lambda\right)^{1+\alpha}\,,
\ee
which reduces to the above identification 
in the limit $\alpha\to0$.
With this normalization, the generalized Chaplygin 
solid Lagrangian can be written as
\be
\mKCh=\rho_{0}\Omega_{\text{DM}} \sqrt{b} \left[
1+ \hat{f}(Y,Z)\left(\frac{\Omega_{\Lambda}}{\Omega_{\text{DM}}}b^{(2\epsilon_0/3-1)/2}\right)^{1+\alpha}\right]^{1/(1+\alpha)}\,,
\label{calK2}
\ee
where $\hat{f}(Y,Z)\equiv f(Y,Z)/f(1/3,1/9)$ 
is normalized such that
$\hat{f}(1/3,1/9)=1$. Let us emphasize that the introduction of the equivalent density parameters does not imply a physical decomposition into DM and DE. As explained above, it merely provides a
parametrization in terms of the corresponding 
$\Lambda$CDM model to which the solid model 
interpolates (see Fig.~\ref{Fig:Chpdensities}).

The transition from the fluid phase to the solid phase can be characterized by the following order parameter:
\begin{equation}
r_f\equiv \left(\frac{\Omega_{\Lambda}}{\Omega_{\text{DM}}}b^{(2\epsilon_0/3-1)/2}\right)^{1+\alpha}=\left(\frac{\Omega_{\Lambda}}{\Omega_{\text{DM}}}a^{(3-2\epsilon_0)}\right)^{1+\alpha}\,,
\label{eq:defrf}
\end{equation}
so the fluid phase occurs for $r_f\ll1$, whereas the solid phase corresponds to $r_f\gg1$. The order parameter evolves as 
$r_f \propto a^{(3-2\epsilon_0)(1+\alpha)}$.
Since $\epsilon_0 \ll 1$ to realize accelerated expansion at late times, this scaling is approximately
$r_f\propto a^{3(1+\alpha)}$. Thus, $r_f$ increases for $\alpha>-1$. We will assume 
\be
\alpha>-1\,,
\ee
in the following, so as to ensure a transition from a fluid phase at early times to a solid phase at late times. This naturally realizes the solid unification of the dark sector: the term $m\sqrt{b}$, which describes a pressureless perfect fluid, can dominate at early times, while the growth of $r_f$ eventually allows the solid phase to take over the evolution and drive a constant-roll quasi-de Sitter phase. 
The evolution of the energy densities for the different species is shown in the left panel of Fig.~\ref{Fig:Chpdensities}. The onset of the solid phase supporting accelerated expansion can be defined as the time $a_\star$ at which the order parameter becomes of order unity, $r_f(a_\star)=1$.\footnote{Although we define the transition by the condition $r_f=1$, the time at which solidification starts to become relevant can be estimated by expanding the energy density for small $r_f$ as $\rho=\mKCh\simeq \rho_{0}\Omega_{\text{DM}} \sqrt{b} 
\left[ 1+r_f/(1+\alpha) \right]$. This shows that the solid contribution starts to become relevant when $r_f\simeq 1+\alpha$.} Since $\epsilon_0$ is assumed to be small, the transition time is simply
\be
a_\star\simeq \left(\frac{\Omega_{\text{DM}}}{\Omega_{\Lambda}}\right)^{1/3}.
\ee

The dark sector and effective equations of state can be written as
\ba
w_{\rm D} &=& -\frac{(3-2\epsilon_0)r_f}
{3(1+r_f)}\,,\label{wD2}\\
w_{\rm eff} &=& 
\frac{\Omega_{\rm r}}{3}
-\frac{(3-2\epsilon_0)
r_f}{3(1+r_f)}\Omega_{\rm D}
=\frac{\Omega_{\rm r}}{3}
+w_{\rm D}\Omega_{\rm D}\,.
\label{weff2}
\ea
At early times, $r_f\ll1$, and hence
$w_{\rm D}\simeq -(1-2\epsilon_0/3)r_f\ll1$,
which is consistent with CDM-like evolution. 
At late times, in the solid phase with $r_f\gg1$, 
the equation of state approaches
$w_{\rm D}\simeq -(1-2\epsilon_0/3)$, corresponding to a constant-roll quasi-de Sitter evolution with $\epsilon=\epsilon_0$.
In the right panel of Fig.~\ref{Fig:Chpdensities}, 
we plot the evolution of $w_{\rm D}$ as a function of $a$ 
for $\epsilon_0=0.01$ and $\alpha=0.05$.
The dark-sector equation of state evolves from the 
early-time value $w_{\rm D}\simeq 0$ to the asymptotic value
$w_{\rm D}\simeq -0.993$. For larger values of $\alpha$,
the transition becomes sharper, and the variation of $w_{\rm D}$ occurs more rapidly.

It is useful to show how the generalized Chaplygin solid described by Eq.~\eqref{calK} is related to the equation of state of the Chaplygin gas model. For this function, the pressure $P_{\rm D}=2b{\cal K}_{,b}-{\cal K}$ is related to the density $\rho_{\rm D}={\cal K}$ as
\be
P_{\rm D}=-\frac{1}{3}
(3-2\epsilon_0)f b^{\epsilon_0 
(1+\alpha)/3} \rho_{\rm D}^{-\alpha}\,.
\ee
In the limit $\epsilon_0 \to 0$, this reduces to $P_{\rm D}=-f \rho_{\rm D}^{-\alpha}$, 
which coincides with the equation of state of the generalized Chaplygin gas model \cite{Bento:2002ps}.  In the same limit, the original Chaplygin gas model \cite{Kamenshchik:2001cp} is recovered as the special case $\alpha=1$. 
Thus, at the background level, the model can reproduce these perfect-fluid 
unified models of DM and DE proposed in the literature. At the level of perturbations, however, our model predicts a different evolution of the cosmic growth history compared to the (generalized) Chaplygin gas models.
This difference arises from the explicit dependence of the function $f$ on $Y$ and $Z$.

%%%%%%%%%%%%%%%%%%%%%%%%%%%%%%
\section{Stability conditions}
\label{sec:stability}
%%%%%%%%%%%%%%%%%%%%%%%%%%%%%%

After showing that the solid unification of the dark sector can generate the desired background evolution, we turn our attention to perturbations, where the distinctive signatures of the solid phase arise. We begin our analysis of cosmological perturbations by deriving the stability conditions associated with the absence of ghosts and Laplacian instabilities. To this end, we compute the quadratic-order action for the perturbations. To simplify the calculation, we focus on the regime dominated by the dark solid and neglect the contribution from the matter sector. This is justified because the matter sector mixes with the dark sector only through gravity. The decoupling of gravity on small scales ensures that the stability conditions derived in this way remain valid when additional minimally coupled matter components, with no direct interactions with the dark sector, are included. In the following, we spare the reader the details of the computation, which are presented in the Appendix with the perturbations from the matter sector taken into account.

We perturb the background configuration \eqref{phiaback} as
\be
\phi^a=\lambda(x^a+\pi^a),
\ee
where $\pi^a$ describes the phonon perturbations of the dark medium.
Although $\pi^a$ carries an internal index, the preserved background
SO(3) symmetry allows us to decompose it into transverse and longitudinal
parts as
\be
\pi^a=\pi^a_T+\partial^a \pi_L\,,
\ee
where $\pi^a_T$ satisfies $\partial_a \pi^a_T=0$. 
The transverse and longitudinal components correspond 
to vector and scalar perturbations, respectively, 
under spatial rotations. In expressions involving internal
indices, we use the Euclidean metric to raise and lower 
indices, e.g.,
$\partial_a \pi_L=\delta_{ab}\partial^b \pi_L$.

For the metric, we use the Newtonian gauge, in which the perturbed line element is parametrized as
\be
\dd s^2=-\big(1+2\psi\big)\dd t^2
+2aS_i\dd x^i\dd t
+a^2\Big[\big(1-2\phi\big)\delta_{ij}+h_{ij}\Big]
\dd x^i\dd x^j\,.
\label{perlineN}
\ee
Here $\psi$ and $\phi$ are the scalar Newtonian gravitational potentials, $S_i$ is a transverse vector perturbation satisfying $\partial_i S^i=0$, and $h_{ij}$ is a transverse-traceless 
tensor perturbation satisfying
$\partial^i h_{ij}=h^i{}_i=0$. 
The latter describes the radiative degrees
of freedom of GWs.

We will eventually decompose the perturbations 
into Fourier modes. For an arbitrary perturbation 
${\cal X}$ in real space, we define
\be
{\cal X}(t,{\bm x})=\int \frac{{\rm d}^3 k}{(2\pi)^3} 
e^{i {\bm k} \cdot {\bm x}} {\cal X}_k (t)\,,
\label{calX}
\ee
where ${\bm k}$ is the comoving wave vector and $k=|{\bm k}|$. 
To simplify the notation, we will omit the 
Fourier-mode label ``$k$''
in what follows whenever no confusion can arise. 
Having introduced the perturbation variables, we
now study each sector separately.

\subsection{Tensor modes}

We begin with the simplest sector, 
namely tensor perturbations. Expanding
the quantities $b$, $Y$, and $Z$ up to quadratic order in $h_{ij}$, we find
\be
b\simeq\frac{\lambda^6}{a^6}
\left(1+\frac{1}{2}h_{ij}h^{ij}\right)\,,\qquad
Y=\frac{1}{3}\left(1+\frac{1}{3}h_{ij}h^{ij}\right)\,,\qquad
Z=\frac{1}{9}\left(1+h_{ij}h^{ij}\right)\,.
\ee
Substituting these expressions into the action \eqref{action}, with
$\Lag_{\text{M}}=0$, gives the corresponding quadratic action. After
performing integrations by parts and using the background equation of motion \eqref{back2}, the quadratic action for tensor perturbations can be written as
\begin{equation}
\mathcal{S}_t^{(2)} = \frac{1}{64\pi G} \int \dd t\,\dd^3x\, a^3
\left[ \dot{h}_{ij}^2 - \frac{(\partial h_{ij})^2}{a^2} 
- \frac{64 \pi G}{9} (\mathcal{K}_{,Y} + \mathcal{K}_{,Z}) h_{ij}^2 \right] \,.
\label{St}
\end{equation}

It is worth mentioning that the quadratic action obtained for GWs is not modified by the addition of other matter components in the form of perfect fluids, since they do not contribute to the tensor sector at linear order. This is corroborated by the fact that the resulting action does not contain derivatives of ${\cal K}$ with respect to $b$, which would survive in the perfect-fluid limit. The absence of ghosts and Laplacian instabilities follows directly from the Einstein-Hilbert term and the absence of nonminimal couplings, which also guarantees exactly luminal propagation for GWs.

On the other hand, although the solid dark sector does not contain tensor perturbations of its own, namely tensor phonons, GWs are affected by the solid through the generation of a time-dependent effective mass determined by $\mK_{,Y}+\mK_{,Z}$. To avoid tachyonic instabilities, we need to require
\be
\mathcal{K}_{,Y}+\mathcal{K}_{,Z} \geq 0\,.
\ee
This is a characteristic feature of cosmological solids with shear rigidity, already encountered in elastic and solid inflation models \cite{Gruzinov:2004ty,Endlich:2012pz}, and arises from the nontrivial realization of an isotropic and homogeneous cosmological background in our setup. This effective mass affects the propagation of GWs through the dark cosmic medium and therefore provides a possible probe of the solid properties of the dark sector. We will explore this possibility in more detail in Sec.~\ref{secGWD}.

\subsection{Vector modes}

We now turn to the vector sector. The vector perturbation $S_i$ is nondynamical and can therefore be integrated out. After doing so, and after performing integrations by parts and using the background equations, the resulting quadratic action for 
$\vec{\pi}_T=\pi_T^a$ in Fourier space reads
\be
{\cal S}_v^{(2)}=\int \frac{{\rm d}t {\rm d}^3 k}{(2\pi)^3}
a^5\frac{b {\cal K}_{,b}}{1+4\epsilon_b a^2H^2/k^2}
\left(\vert\dot{\vec{\pi}}_T\vert^2-c_T^2\frac{k^2}{a^2} \vert\vec{\pi}_T\vert^2\right)\,,
\label{Svform}
\ee
where we have defined
\be
\epsilon_b = \frac{8 \pi G b {\cal K}_{,b}}{H^2}\,,\qquad 
c_T^2=\bar{c}_T^2\left(1+4\epsilon_b \frac{a^2H^2}{k^2}\right)\,,
\qquad\text{with}\qquad \bar{c}_T^2= \frac{2}{9}\frac{{\cal K}_{,Y}+{\cal K}_{,Z}}
{b {\cal K}_{,b}}\,.
\label{cTdef}
\ee

In the regime where the dark solid dominates, we have $\epsilon_b=\epsilon$.
However, we keep $\epsilon_b$ for generality, since the above expression
also applies when the dark sector is not the dominant component. The absence
of ghosts is guaranteed for
\be
b\mK_{,b}>0\,.
\label{NGvec}
\ee
This amounts to imposing the null energy condition
$\rho_\text{D}+p_{\text{D}}>0$, 
as in the case of a perfect fluid. Thus,
the ghost-free condition is insensitive to the solid properties.

On the other hand, the absence of Laplacian instabilities, together with the absence of ghosts, requires
\be
\mK_{,Y}+\mK_{,Z}>0\,.
\label{Lapvec}
\ee
Under this condition, tachyonic instabilities for GWs are also absent. In
the perfect-fluid limit, where $\mK_{,Y}+\mK_{,Z}=0$, we recover the usual
result $c_T^2=0$, namely, there are no transverse phonon waves. This is the expected behavior in our unified scenario at high redshifts, where the dark solid behaves approximately as a pressureless fluid. At late times, when
the dark sector enters the solid phase, 
transverse phonons become genuine propagating modes and may therefore affect the vector perturbation sector.

\subsection{Scalar modes}

Let us finally consider the most involved sector, 
namely scalar perturbations. In this case, 
the perturbation $\psi$ is nondynamical and can therefore be
integrated out. After doing so, performing integrations by parts, and using the background equations, the perturbation $\phi$ drops out. We are then left with the quadratic action for the only scalar dynamical degree of freedom, corresponding to the longitudinal phonon $\pi_L$. The resulting quadratic action can be written in Fourier space as
\be
\mS_{L}^{(2)}=\frac{1}{2}\int\frac{{\rm d}t {\rm d}^3 k}{(2\pi)^3}a^5\frac{ k^2 b\mK_{,b}}{1+3\epsilon_b 
a^2H^2/k^2}\left[\vert\dot{\pi}_L\vert^2 -\left(\frac{c_L^2k^2}{a^2}+m_L^2\right)\vert\pi_L\vert^2\right]\,,
\label{eq:S2L}
\ee
where the propagation speed squared 
and the effective mass squared are given by
\begin{equation}
c_L^2=\frac{\bar{c}_L^2}{1+3\epsilon_b a^2H^2/k^2},\qquad
m^2_L=-\frac{\epsilon_b H^2}{1+3\epsilon_b a^2H^2/k^2}
\left[1-8\bar{c}_T^2\left(1+\frac{3\epsilon_b}{2}\frac{a^2H^2}{k^2}\right)\right],
\end{equation}
with
\be
\bar{c}_L^2\equiv 1+2\frac{b\mK_{,bb}}{\mK_{,b}}+\frac{8}{27}\frac{\mK_{,Y}+\mK_{,Z}}{b\mK_{,b}}.
\ee

The absence of ghost instabilities again requires $b\mK_{,b}>0$, 
as for the transverse phonons, and therefore does not lead to any new condition. 
Regarding Laplacian instabilities, we need $c_L^2>0$, which,
together with the no-ghost condition, imposes $\bar{c}_L^2>0$. 
In the Appendix, we show that the same linear stability conditions remain valid even in the presence of matter fluids, including baryons and radiation.
On small scales, $k\gg aH$, the propagation speed reduces to
$c_L^2\simeq\bar{c}_L^2$, which represents the propagation 
speed squared of longitudinal phonons on sub-Hubble scales. Comparing this expression with that of $\bar{c}_T^2$, 
we obtain the relation\footnote{In the solid
inflation scenario \cite{Endlich:2012pz}, requiring inflation to last for a sufficient number of e-folds leads to the relation
$\bar{c}_T^2\simeq (3/4)\big(1+\bar{c}_L^2\big)$, 
which imposes an additional constraint on the 
propagation speeds. Our unified solid dark
sector is not subject to this constraint, since we 
do not need accelerated expansion to last for many e-folds. 
Rather, it is sufficient to have a short period of acceleration, 
roughly during the last e-fold of the cosmic evolution.}
\be
\bar{c}_L^2
=1+2\frac{b\mK_{,bb}}{\mK_{,b}}
+\frac{4}{3}\bar{c}_T^2\,.
\ee
This expression can be written in the 
more suggestive form
\be
\bar{c}_L^2
=\frac{\partial_b\PD}{\partial_b\rhoD}
+\frac{4}{3}\bar{c}_T^2\,,
\ee
where the first term, $\partial_b\PD/\partial_b\rhoD$, 
is the propagation speed squared in the perfect-fluid limit. Equivalently, this term can be written as
$\partial P_D/\partial\rho_D|_{Y,Z}$, 
corresponding to the usual adiabatic
sound speed squared. This form shows that the 
additional contribution proportional to $\bar{c}_T^2$ 
originates from the solid properties of the dark medium.

In our unified scenario, where $\mK\simeq m\sqrt{b}$ 
at high redshifts, we have 
$\partial_b\PD/\partial_b\rhoD\simeq 0$ and 
hence $\bar{c}_L^2\simeq 4\bar{c}_T^2/3\ll1$, 
consistently with the fact that the dark sector behaves approximately as a pressureless fluid. 
It is also interesting to note that an accelerated regime characterized by a constant slow-roll parameter 
$\epsilon \ll 1$, for which $\mK\propto b^{\epsilon/3}$,
gives $\partial_b\PD/\partial_b\rhoD\simeq -1$. 
In this case, the longitudinal propagation speed 
squared becomes
\be
\bar{c}_L^2\simeq -1+\frac{4}{3}\bar{c}_T^2\,.
\label{eq:cLepsilon0}
\ee
This shows that the solid contribution helps cure the Laplacian instability that would be present in the pure-fluid regime. Thus, solid deformations can stabilize the accelerated regime and widen the range of viable accelerated cosmologies.

The above relation also illustrates why a fluid unification 
of the dark sector runs into difficulties. In that case, $\bar{c}_T^2=0$, and the phase
of accelerated expansion has
$\bar{c}_L^2=1+2b\mK_{,bb}/\mK_{,b}$. 
This is the origin of the prominent
acoustic oscillations, or Laplacian instabilities, of the perturbations reported in Ref.~\cite{Sandvik:2002jz}. 
In the solid unification, by contrast, 
the additional contribution from $\bar{c}_T^2$ 
can tame these undesired features.

Interestingly, the solid properties are governed by the single combination
$\mK_{,Y}+\mK_{,Z}$. Therefore, the same parameter that stabilizes scalar
perturbations and suppresses the undesired fluid-like behavior also generates
distinctive observational signatures that can serve as smoking guns of the
solid nature of the dark sector. This is the main advantage of the solid
unification over the fluid case. In the latter, it was argued in
Ref.~\cite{Sandvik:2002jz} that realistic unified fluid models need to be
very close to $\Lambda$CDM and, in practice, become almost indistinguishable
from it. As we will see, the solid unification can instead generate
distinctive signatures.

The function $m_L^2$ describes the effective mass squared of the phonons.
Unlike ghost or Laplacian instabilities, a negative value of $m_L^2$ does
not necessarily signal a pathological instability. Rather, it can correspond
to the standard Jeans-type instability responsible for the gravitational
growth of structures. For sub-Hubble modes, the effective mass squared
reduces to
\be
m_L^2(k\gg aH)\simeq
-\left(1-8\bar{c}_T^2\right)\epsilon_b H^2\,.
\ee
Deep in the matter era, when the dark sector is in the fluid-like regime and baryons and radiation are neglected, 
we approximately have $2b {\cal K}_{,b} \simeq {\cal K}$,
$3H^2 \simeq 8\pi G {\cal K}$, $\epsilon_b \simeq 3/2$, and
$\bar{c}_T^2\ll1$. We then recover the usual result
$m_L^2\simeq -3H^2/2$, corresponding to the Jeans instability of pressureless DM.
This negative effective mass squared is precisely what allows density perturbations to grow gravitationally. 
As the dark sector enters the solid phase, $\bar{c}_T^2$ 
grows and $\epsilon_b$ becomes very small. Both effects
reduce the magnitude of the effective mass squared, so weaker growth is expected. It is important to keep in mind, however, that the effective sound horizon of these perturbations can also be modified and must be taken into account when analyzing the evolution of $\pi_L$. We will perform a more
detailed analysis of structure growth in the following sections.

Having obtained the theoretical viability conditions from the absence of fatal instabilities, we now analyze 
the evolution of scalar perturbations
and the possible signatures of the solid phase.

%%%%%%%%%%%%%%%%%%%%%%%%%%%%%%%%%%%%
\section{Gauge-invariant scalar perturbations}
\label{perturbsec}
%%%%%%%%%%%%%%%%%%%%%%%%%%%%%%%%%%%%

In this section, we derive the scalar perturbation equations of motion by including the contributions arising from the matter Lagrangian ${\cal L}_{\rm M}$, namely those of baryons and radiation. We would like to clarify how the solid properties of the unified dark sector make it possible to evade the observational constraints found in Ref.~\cite{Sandvik:2002jz} for fluid unification.

\subsection{Gauge invariant variables}

In the following, we do not specify to any particular 
gauge for scalar perturbations, 
but instead work with gauge-invariant variables. 
For the gravitational sector, we consider 
the perturbed line element
\begin{equation}
{\rm d}s^2 = -(1 + 2A){\rm d}t^2 
+ 2a \partial_{i}\chi{\rm d}t {\rm d}x^i
+ a^2(t) \Big[ (1 + 2\zeta) \delta_{ij}
+ 2 \partial_i \partial_j E  \Big]
{\rm d}x^i {\rm d}x^j \,,
\label{perline}
\end{equation}
where $A$, $\chi$, $\zeta$, and $E$ describe scalar perturbations.
Under an infinitesimal gauge transformation
$\tilde{x}^\mu=x^\mu+\xi^\mu$, where
$\xi^\mu=(\xi^0,\partial^i\xi)$ for scalar modes, 
the metric perturbations
and the longitudinal phonon perturbation transform as
\begin{equation}
\tilde{A} = A - \dot{\xi}^0\,, \qquad 
\tilde{\chi} = \chi + \frac{\xi^0}{a} - a \dot{\xi}\,, \qquad 
\tilde{\zeta} = \zeta - H \xi^{0}\,, \qquad 
\tilde{E} = E - \xi\,, \qquad 
\tilde{\pi}_L = \pi_L - \xi\,.
\end{equation}
By virtue of the above transformation properties, we can construct the following gauge-invariant quantities in the scalar sector:
\begin{equation} 
\Psi \equiv A + \frac{{\rm d}}{{\rm d}t} \left[ a (\chi - a \dot{E}) \right]\,, \qquad
\Phi \equiv -\zeta - aH \left( \chi - a \dot{E} \right)\,,\qquad 
\Pi_L \equiv \pi_L - E\,.
\label{Grapo}
\end{equation}

We will also consider a perturbed matter sector 
described by an energy-momentum tensor of the form
\be
T_{(i)}^{\mu\nu}=\Big(\rho_{(i)}+P_{(i)}\Big) u_{(i)}^\mu u_{(i)}^\nu+P_{(i)}g^{\mu\nu}\,,
\ee
where $\rho_{(i)}$ and $P_{(i)}$ are the background energy density and pressure of the $i$-th component, and $u^\mu_{(i)}$ 
is its four-velocity. 
At the perturbative level, this matter sector is 
described by the corresponding perturbations of the energy density $\delta\rho_{(i)}$ and pressure $\delta P_{(i)}$, as well as by the peculiar velocity $\vec{u}{(i)}$, which can be decomposed into transverse and longitudinal components as $\vec{u}_{(i)}=\vec{v}_{(i)}+\nabla v_{(i)}$, with $\nabla\cdot\vec{v}_{(i)}=0$ and $v_{(i)}$ being the velocity scalar potential. In this work, we do not consider possible anisotropic stresses in this sector, although they can be straightforwardly incorporated.
The perturbed energy densities and velocity potential of the fluid transform as 
\be
\delta\tilde{\rho}_{(i)}=\delta \rho_{(i)}
-\dot{\rho}_{(i)}\xi^0,
\qquad \tilde{v}_{(i)}=v_{(i)}-\xi^0\,,
\qquad 
\delta\tilde{P}_{(i)}=\delta P_{(i)}
-\dot{P}_{(i)}\xi^0\,.
\ee
This allows us to define the following 
gauge-invariant quantities in the matter sector:
\begin{equation}
\delta \rho_{(i),{\rm N}}= \delta \rho_{(i)} 
+ a \dot{\rho}_{(i)} \left( \chi - a \dot{E} \right), \qquad
v_{(i),{\rm N}} = v_{(i)} + a \left( \chi - a \dot{E} \right)\,,
\qquad
\delta P_{(i),{\rm N}}=\delta P_{(i)}
+a \dot{P}_{(i)} \left( \chi - a \dot{E} \right)\,.
\label{deltarho}
\end{equation}
We also introduce the corresponding 
total quantities: 
\be
\delta \rho_{{\rm N}} \equiv 
\sum_{i}\delta \rho_{(i),{\rm N}}\,,
\qquad 
(\rho_{\rm M}+P_{\rm M})v_{\rm N}\equiv 
\sum_{i} (\rho_{(i)}+P_{(i)})v_{(i),{\rm N}}\,,
\qquad
\delta P_{{\rm N}} \equiv 
\sum_{i}\delta P_{(i),{\rm N}}\,,
\ee
where $\rho_{\rm M}=\sum_{i}\rho_{(i)}$ and 
$P_{\rm M}=\sum_{i}P_{(i)}$.
In the following, we focus on scalar perturbations and work with the gauge-invariant quantities $\Psi$, $\Phi$, and $\Pi_L$, which are directly
related to the perturbation variables in the Newtonian gauge.

\subsection{Gravitational equations: 
A solid contribution to the slip parameter}

The gravitational perturbation equations can be obtained 
by perturbing the Einstein equations in the usual way,
\begin{equation}
\delta G^\mu{}_\nu=8\pi G\delta T^\mu{}_\nu\,,
\end{equation}
where $\delta G^\mu{}_\nu$ is the perturbed Einstein tensor. 
The perturbed energy-momentum tensor $\delta T^\mu{}_\nu$ 
contains contributions from the solid dark sector 
and from all other species. 
In Fourier space, the scalar perturbation 
equations can then be written as 
\begin{eqnarray}
3H \left( \dot{\Phi}+H \Psi \right)
+\frac{k^2}{a^2}\Phi
&=&-4\pi G \left( \delta \rho_{\rm D}
+\delta \rho_{\rm N} \right)\,,
\label{Eq:dG00}\\
\dot{\Phi}+H \Psi&=&4\pi G 
\left[ (\rho_{\text{D}}+P_{\text{D}}) v_{\text{D}}
+(\rho_{\rm M}+P_{\rm M}) v_{\rm N}\right] 
\,,\label{Eq:dG0i}\\
\ddot\Phi+H(3\dot{\Phi}+\dot{\Psi})
+\big(3H^2+2\dot{H}\big)\Psi+\frac{k^2}{3a^2}\big(\Phi-\Psi\big)&=&4\pi G \left( \delta P_{\rm D}
+\delta P_{\rm N} \right)\, ,\label{Eq:dGTr}\\
\Psi-\Phi&=&-\frac{64\pi G a^2}{9} 
\big( {\cal K}_{,Y}+ {\cal K}_{,Z} \big)\Pi_L\,,
\label{Eq:dGan}
\end{eqnarray}
where 
\be
\delta\rho_{\text{D}}=-2b\mK_{,b}
\big(k^2\Pi_L-3\Phi\big),\qquad
(\rho_{\text{D}}+P_{\text{D}}) v_{\text{D}}=2a^2b\mK_{,b} \dot{\Pi}_L\,,\qquad
\delta P_{\text{D}}=2b\mK_{,b}\left(1+\frac{2b\mK_{,bb}}{\mK_{,b}}\right)\Big(3\Phi-k^2\Pi_L\Big)\,.
\label{eq:DSperturbations}
\ee
From the last expression, we can read off 
the velocity potential associated 
with the dark sector as
\be
v_{\text{D}}=a^2\dot{\Pi}_L\,.
\ee
The gravitational equations already display a distinctive signature of the
solid dark sector through the shear equation \eqref{Eq:dGan}, where the
anisotropic stress\footnote{Here we neglect possible contributions to the
anisotropic shear from standard components, such as neutrinos. 
We assume that these contributions are negligible compared to 
the solid anisotropic stress.}
of the solid generates a nontrivial slip parameter,
\be
\gamma \equiv 
\frac{\Phi}{\Psi} = 1+\frac{64\pi G a^2}{9} 
\left( {\cal K}_{,Y}+ {\cal K}_{,Z} \right)
\frac{\Pi_L}{\Psi}\,.
\ee
This expression clearly shows how the solidity of the unified dark sector generates a deviation from $\gamma=1$, driven by the combination $\mK_{,Y}+\mK_{,Z}$. 
As discussed above, this is the same combination that
controls the distinctive signatures of the solid phase. The slip parameter can also be expressed in the more suggestive form
\be
\gamma=1+16\pi G\,\bar{c}_T^2\big(\rhoD+\PD\big)
\frac{\Pi_L}{\Psi}\,,
\ee
where we have used the definition of $\bar{c}_T^2$ 
in Eq.~\eqref{cTdef} and
the relation $2b\mK_{,b}=\rho_{\rm D}+P_{\rm D}$. 
At high redshifts, our unified solid dark sector satisfies $\bar{c}_T^2\ll1$ by construction, so $\gamma\simeq1$ 
at early times. As the dark sector enters the solid phase,
$\bar{c}_T^2$ becomes increasingly more relevant and a nontrivial slip parameter can be generated. 
It is worth noting that this scenario provides an example in which an anomalous slip parameter arises not from a modification of gravity, but from the anisotropic stress of the solid dark sector.

We can obtain a more informative expression for the slip parameter by deriving a relation between $\Pi_L$ and $\Psi$. To this end, we first note
that $\Phi$ can be expressed in terms of the matter perturbations and the
solid dark-sector phonon $\Pi_L$ by combining the Hamiltonian and momentum
constraint equations \eqref{Eq:dG00} and \eqref{Eq:dG0i}. 
The resulting expression is
\be
\Phi = \frac{4\pi G a^2\Big[2b {\cal K}_{,b} 
(k^2 \Pi_L-3a^2 H \dot{\Pi}_L)-\delta \rho_{\rm N}
-3H (\rho_{\rm M}+P_{\rm M})v_{\rm N}\Big]}
{k^2+24\pi G a^2 b {\cal K}_{,b}}\,.
\label{Phieq}
\ee
At late times, when the solid dark sector dominates, we can safely neglect
the contribution from the additional matter fields, 
since baryon and photon perturbations do not contribute appreciably. In that regime, the relation
between the gravitational potential and the phonon field becomes
\be
\Phi = \frac{\epsilon\,a^2H^2}{1+3\epsilon a^2 H^2/k^2}
\left(\Pi_L-\frac{3a^2 H^2}{k^2}\frac{\dot{\Pi}_L}{H}\right).
\label{Phieq2}
\ee
where we have used $8\pi G b\mK_{,b}=-\dot{H}=\epsilon H^2$. 
Taking the small-scale limit $k\gg aH$, we obtain
\be
\Phi \simeq \epsilon a^2H^2\Pi_L\,,
\ee
where we have neglected $H\dot{\Pi}_L$ relative to $k^2\Pi_L$. This is
justified for slow modes with $\dot{\Pi}_L\sim H\Pi_L$ 
that do not oscillate, although it can break down for rapidly oscillating modes. Even in that case, the estimate remains parametrically useful. 
The potential $\Psi$ can then be
obtained from the shear equation \eqref{Eq:dGan} as
\begin{eqnarray}
\Psi&=&\Phi-\frac{64\pi G a^2}{9}(\mK_{,Y}+\mK_{,Z})\Pi_L\nonumber\\
&=&\Phi-4\bar{c}_T^2\epsilon a^2 H^2\Pi_L\,,
\label{Psieq}
\end{eqnarray}
so that, in the regime of dark-sector domination and 
on sub-Hubble scales, the slip parameter can be approximated as
\be
\gamma \simeq
\frac{1}{1-4\bar{c}_T^2}\,.
\ee
This expression shows that the slip parameter is scale independent within
the small-scale approximation adopted here, although it generally evolves in
time through the evolution of $\bar{c}_T^2$. As explained above,
$\bar{c}_T^2\ll1$ at high redshifts and becomes relevant only in the
late-time solid phase. 
Since $\bar{c}_T^2$ is the quantity that helps
stabilize the scalar sector, there is a clear link 
between the stabilization of scalar perturbations and the generation of a nontrivial slip parameter.

It is interesting to note that some studies have reported 
possible deviations from the standard value $\gamma=1$, 
suggesting the presence of nonstandard
anisotropic stress \cite{Arjona:2020kco,Alestas:2022gcg}
(see also \cite{Pinho:2018unz,Skara:2019usd,Yang:2020wby}). A
scale-dependent reconstruction was also performed in Ref.~\cite{Guerrini:2023pre},
where the cosmological slip was found to remain essentially unconstrained.
Since our slip parameter becomes scale independent on small scales,
constraints on the slip parameter from those scales may translate into
constraints on the unified solid dark-sector model. It is important, however,
to ensure that the relevant scales remain within the regime of validity of
the model, since nonlinear evolution may become important and higher-order
operators may also become relevant. For instance, applying Solar-System
constraints would require resolving the density profile of the galactic halo.

\subsection{Solid growth}

The gravitational equations are supplemented 
with the equation for the phonon field
\be
\ddot{\Pi}_L-H \left(1+\frac{6b{\cal K}_{,bb}}{{\cal K}_{,b}} \right)
\dot{\Pi}_L+\bar{c}_L^2 \frac{k^2}{a^2}\Pi_L=
\frac{1}{a^2} \left[ \Psi+3 \left(1+\frac{2b{\cal K}_{,bb}}
{{\cal K}_{,b}} 
\right) \Phi\right]\,,
\label{Eq:PiL}
\ee
as well as by the continuity and Euler equations for each component of the matter sector,
\ba
& &
\delta \dot{\rho}_{(i),{\rm N}}+3 \big(1+c_{(i)}^2\big)H 
\delta \rho_{(i),{\rm N}}+(\rho_{(i)}+P_{(i)})
\left( \frac{k^2}{a^2} v_{(i),{\rm N}}-3\dot{\Phi} 
\right)=0\,,\label{pereq5D}\\
& &
\dot{v}_{(i),{\rm N}}-3H c_{(i)}^2 v_{(i),{\rm N}}
-\Psi-\frac{c_{(i)}^2}{\rho_{(i)}+P_{(i)}} 
\delta \rho_{(i),{\rm N}}=0\,,\label{pereq6D}
\ea
where $c_{(i)}^2=\dot{P}_{(i)}/\dot{\rho}_{(i)}$. 
We assume that the different components 
labeled by $i$ interact only gravitationally.

The dynamics of the longitudinal phonon $\Pi_L$ is affected by the gravitational potentials appearing on the right-hand side of Eq.~\eqref{Eq:PiL}. In the presence of other matter fields, these potentials will be sourced by all the components as shown by Eq.~\eqref{Phieq}. That equation, together with the shear equation \eqref{Eq:dGan} allow to remove the gravitational potentials from the phonon equation and express it as
\be
\ddot{\Pi}_L+H\left(3-\frac{1+3\bar{c}_L^2-4\bar{c}_T^2}
{1+3\epsilon_b a^2H^2/k^2}\right)\dot{\Pi}_L+\left( c_L^2 \frac{k^2}{a^2}+m_L^2\right)\Pi_L=
\frac{8 \pi G (2{\cal K}_{,b}+3b {\cal K}_{,bb})}
{{\cal K}_{,b}(k^2+24\pi G a^2 b {\cal K}_{,b})}
\Big[ \delta \rho_{\rm N}+3H (\rho_{\rm M}
+P_{\rm M})v_{\rm N} \Big]\,.
\label{pieq}
\ee
For modes well-inside the Hubble horizon, we have $c_L^2\simeq\bar{c}_L^2$ and $m_L^2 \simeq -(1 - 8\bar{c}_T^2)\epsilon_b H^2$, so the equation for the longitudinal phonons reduces to
\be
\ddot{\Pi}_L+H\left(2-3\bar{c}_L^2-4\bar{c}_T^2\right)\dot{\Pi}_L+\left[ \bar{c}_L^2 \frac{k^2}{a^2}-(1-8\bar{c}_T^2)\epsilon_bH^2\right]\Pi_L=\frac{16 \pi G}{k^2} \left(1+\frac{3b \mK_{,bb}}{2\mK_{,b}}\right)
\Big[ \delta \rho_{\rm N}+3H (\rho_{\rm M}+P_{\rm M})
v_{\rm N} \Big]\,.
\label{pieqsub}
\ee
As long as $2{\cal K}_{,b} + 3b {\cal K}_{,bb} \neq 0$, the presence of baryons and radiation affects the dynamics of $\Pi_L$. This is because the gravitational potentials $\Psi$ and $\Phi$ in the phonon equation contain contributions from the perturbations $\delta \rho_{\rm N}$ and $v_{\rm N}$. 

At high redshifts, the dark sector is in the fluid phase and all solid
contributions are negligible, so the evolution of perturbations is the same
as in $\Lambda$CDM. In particular, prior to radiation-matter equality, the
right-hand side of Eq.~\eqref{pieq} acts as an external source provided by
the radiation component, and the dark-sector density perturbation evolves as
the CDM component in the standard model.

After equality, the energy content of the Universe is dominated by the unified dark sector, so that the right-hand side of the same equation can be neglected. Although radiation rapidly becomes negligible, baryons still contribute a fraction of the total matter sector, and a detailed analysis throughout matter domination should therefore take them into account. For our purposes here, however, we neglect both baryons and radiation. In this case, the source term vanishes, and the longitudinal scalar perturbation $\Pi_L$ obeys a single second-order differential equation. This is precisely the equation of motion corresponding to the quadratic action 
derived from Eq.~\eqref{eq:S2L}.

The sound horizon scale $\kL$ for the longitudinal phonon 
is parametrically given by
\be
\kL^2\equiv\frac{1-8\bar{c}_T^2}{\bar{c}_L^2}\epsilon_b (aH)^2.
\ee
At high redshifts, we have $\bar{c}_L^2,\bar{c}_T^2\ll1$ and
$\epsilon_b\simeq 3/2$, so the sound-horizon scale is pushed to
$\kL\to \infty$. This means that all modes are effectively outside the
sound horizon, and the longitudinal perturbation grows at most as
$\Pi_L\propto a$ after the onset of matter domination, without undergoing
oscillations induced by the pressure term $\bar{c}_L^2 k^2/a^2$. This
behavior should not be modified by the presence of baryons, since baryons
are also pressureless.

To be more explicit, since $\bar{c}_L^2,\bar{c}_T^2\ll1$ at high redshifts,
where the dark sector is in the pressureless fluid regime, the phonon field
evolves, neglecting baryons and radiation, according to
\begin{equation}
\ddot{\Pi}_L+2H\dot{\Pi}_L-\epsilon H^2 \Pi_L \simeq 0\,,
\label{Pima}
\end{equation}
with $\epsilon \simeq 3/2$ and $H \simeq 2/(3t)$. 
This equation can be integrated to give
\begin{equation}
\Pi_L=c_1 t^{2/3}+\frac{c_2}{t}\,,
\end{equation}
where $c_1$ and $c_2$ are constants. Therefore, deep in the matter-dominated era, the longitudinal perturbation grows as $\Pi_L \propto t^{2/3} \propto a$. This is nothing but the standard growth of DM perturbations in the matter dominated epoch. To see how this growing mode directly translates into the growth of density perturbations, let us notice that the density contrast of the solid dark sector can be related to the phonon field and the gravitational 
potential from Eq.~\eqref{eq:DSperturbations} as
\be
\delta_{\rm D}=\frac{\delta \rho_{\rm D}}{\rho_{\rm D}}
=-\frac{2b {\cal K}_{,b}}{{\cal K}} 
\left( k^2 \Pi_L-3\Phi \right)\,.
\label{deltaD}
\ee
In the absence of radiation and baryons, we can use Eq.~\eqref{Phieq} 
to obtain the following expression for the density contrast 
in terms of the phonon field:
\be
\delta_{\rm D}=-\frac{2}{3} \frac{\epsilon\, k^2}
{1+3\epsilon a^2 H^2/k^2} 
\left[ \Pi_L+9 \epsilon \left( \frac{aH}{k}\right)^4 
\frac{\dot{\Pi}_L}{H}\right]\,.
\label{deltaes}
\ee
Deep inside the Hubble horizon and for slow modes with $\dot{\Pi}_L\sim H\,\Pi_L$, this expression reduces to\footnote{This can be obtained in a more direct manner by noticing that, on sub-Hubble scales, the Poisson equation reduces to $\Phi \simeq -(4\pi Ga^2\rho_{\text{D}}/k^2)\delta_{\text{D}}
\simeq -[3a^2H^2/(2k^2)]\delta_{\text{D}}\ll\delta_{\text{D}}$, which means that Eq. \eqref{deltaD} can be approximated to $\delta_{\text{D}}\simeq -(1+w_{\text{D}})k^2\Pi_L$.}
\be
\delta_{\text{D}}\simeq -\frac{2}{3}\epsilon \,k^2\Pi_L.
\label{eq:deltaPi}
\ee
In the pressureless fluid phase of the dark sector, we have
$\epsilon=3/2$, so $\delta_{\rm D}\simeq-k^2\Pi_L$. As anticipated,
the growing mode of $\Pi_L$ directly reproduces the standard CDM growth,
$\delta_{\rm D}\propto a$. In the same regime, we can use the relation
\eqref{eq:deltaPi} in the phonon-field equation \eqref{pieqsub} to obtain
the following equation for the evolution of the density contrast:
\be
\ddot{\delta}_{\rm D}+\left(
2 -3\bar{c}_L^2+4\bar{c}_T^2 
-\frac{2\dot{\epsilon}}{H \epsilon} \right) 
H\dot{\delta}_{\rm D}
-\left( 4\pi G_{\rm eff}\rho_{\rm D}
-\bar{c}_L^2 \frac{k^2}{a^2} 
\right) \delta_{\rm D} \simeq 0\,,
\ee
where we have defined the effective gravitational constant for the dark sector density contrast $\delta_{\text{D}}$ as
\be
G_{\rm eff}=\frac{2}{3}G
\left[ 
\epsilon (1-8 \bar{c}_T^2)
+\frac{\ddot{\epsilon}}{H^2 \epsilon}+\left( 2 -3\bar{c}_L^2+4\bar{c}_T^2\right) \frac{\dot{\epsilon}}{H\epsilon}-\frac{2 \dot{\epsilon}^2}{H^2\epsilon^2}
\right]\,.
\ee

It is important to keep in mind that this quantity acts as an effective
gravitational constant only for the dark-sector density contrast. It does
not imply that the entire structure-formation process can be described by a
single modified gravitational constant. For instance, baryons still respond
to the standard gravitational constant $G$. Deep in the matter-dominated
era, where $\epsilon \simeq 3/2$, we have
$G_{\rm eff}\simeq G(1-8\bar{c}_T^2)$. Initially, $\bar{c}_T^2$ is close to
zero, but its growth at low redshifts leads to a suppression of
$G_{\rm eff}$. In addition, the nonvanishing sound speed squared
$\bar{c}_L^2$ of the longitudinal mode provides pressure support against
gravitational collapse, thereby suppressing the growth of
$\delta_{\rm D}$. This behavior parallels the suppression mechanism
discussed above for the perturbation $\Pi_L$.

\subsection{Effects on baryons}

We now briefly discuss how baryon perturbations are affected by the modified gravitational potentials generated predominantly by dark-sector perturbations at low redshifts. 
Baryon perturbations obey the standard continuity and Euler
equations \eqref{pereq5D} and \eqref{pereq6D} 
with vanishing pressure. Introducing
the baryon density contrast as
$\delta_{\rm b}\equiv\delta\rho_{\rm b}/\rho_{\rm b}$, we obtain
\ba
\dot\delta_{\rm b}
+\frac{k^2}{a^2}v_{\rm b}
-3\dot\Phi &=&0\,,
\label{db-cont}  \\
\dot v_{\rm b} &=&\Psi\,.
\label{vb-euler}
\ea
The potential $\Psi$ determines the motion of nonrelativistic 
tracers, while the time variation of $\Phi$ enters the continuity equation. Taking the time derivative of Eq.~\eqref{db-cont} 
and using Eq.~\eqref{vb-euler}, we find
\be
\ddot\delta_{\rm b}
+2H\dot\delta_{\rm b}
+\frac{k^2}{a^2}\Psi
=3\ddot{\Phi}+6H\dot{\Phi}\,.
\label{db-second}
\ee
On scales well inside the Hubble radius, the time derivatives of $\Phi$
are usually subdominant, so Eq.~(\ref{db-second}) reduces to
\be
\ddot\delta_{\rm b}
+2H\dot\delta_{\rm b}
\simeq -\frac{k^2}{a^2}\Psi\,.
\label{db-qs}
\ee
The potential $\Psi$ is not an independent external source, but is obtained from Eq.~\eqref{Psieq} after solving the perturbation equation for the longitudinal mode $\Pi_L$. 
If the evolution of $\Psi$ is modified by the
solid dark-sector perturbation, this directly affects the growth of
$\delta_{\rm b}$ through Eq.~\eqref{db-qs}. Thus, within the approximation
adopted here, baryons act as nonrelativistic tracers of the gravitational
potential generated predominantly by the dark sector.

%%%%%%%%%%%%%%%%%%%%%%%%%%%%%%%
\section{Observational signatures of a case study: 
The generalized Chaplygin solid}
\label{signaturemodel}
%%%%%%%%%%%%%%%%%%%%%%%%%%%%%%%

In this section, we illustrate the observational signatures of the solid unification of the dark sector for the model introduced in Sec. \ref{sec:GChs}, whose Lagrangian is given by \eqref{calK}. We take this model as a proxy for the general scenario, but it is important to keep in mind that some of the features obtained below are not generic to the whole class of unified solid dark-sector scenarios. This choice is further motivated by the generalized Chaplygin gas analyzed in \cite{Sandvik:2002jz}, since it allows us to clearly identify the effects of solidifying the accelerated regime. We should note that the generalized Chaplygin gas considered in \cite{Sandvik:2002jz} corresponds to $\epsilon_0=0$. However, we will see that this is not a smooth limit of the expressions obtained below, and therefore this limit involves some subtleties that must be taken into account. In any case, this model serves our purpose of illustrating how the solid unification improves upon the fluid unified scenario. 

We will consider a regime in which the solid dark sector dominates. We therefore focus on the evolution 
after radiation-matter equality and neglect the 
contribution from baryons. Our aim is to illustrate how the solid phase can potentially give rise to distinctive observational signatures, even when the background evolution and structure formation are essentially indistinguishable from those in the corresponding $\Lambda$CDM model. For the background evolution, we fix the parameters to $\Omega_{\text{DM}}=0.3$, and $\Omega_{\Lambda}=0.7$.

\subsection{Viability}

\begin{figure}[t!]
\centering
\includegraphics[width=0.49\textwidth]{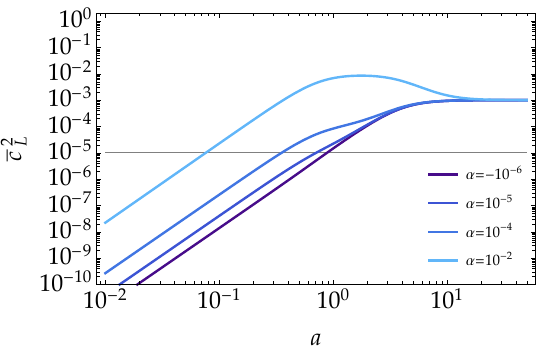}
\includegraphics[width=0.49\textwidth]{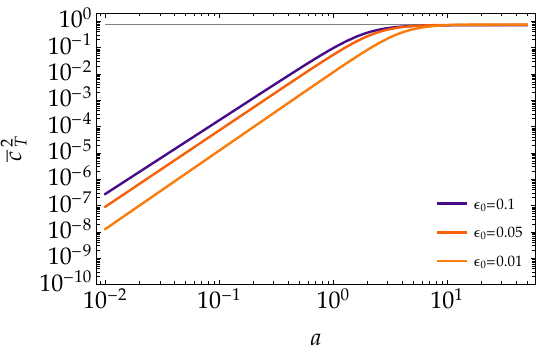}
\caption{Left panel: Evolution of $\bar{c}_L^2$ on sub-Hubble scales for fixed values of $\bar{c}_{L,\text{asy}}^2=10^{-3}$ and $\epsilon_0=0.01$, and for several values of $\alpha$. For the chosen parameters, the propagation speed squared grows at early times approximately as $\bar{c}_L^2\propto a^{3}$ and approaches its constant asymptotic value in the future. The gray line indicates the estimated observational bound that should be satisfied. Right panel: Evolution of $\bar{c}_T^2$ for $\alpha=10^{-5}$ and $\bar{c}_{L,\text{asy}}^2=10^{-3}$, and for different values of $\epsilon_0$. The gray line shows the universal value $3/4$ discussed in the main text.}
\label{fig:cLb}
\end{figure}

For the generalized Chaplygin model described by \eqref{calK}, the squared propagation speeds of the vector and longitudinal modes in the sub-Hubble regime are given, respectively, by
\ba
\bar{c}_T^2 &=& \frac{4(f_{,Y}+f_{,Z})}
{3(1+\alpha)(2\epsilon_0 r_f+3)f}r_f\,,
\label{cV}\\
\bar{c}_L^2 &=& \frac{3(1 +\alpha) 
(2 \epsilon_0-3)
\Big[\alpha \big(2 \epsilon_0 - 3\big) 
+ 2 \epsilon_0 \big(r_f + 1\big)\Big]f 
+ 16\big(r_f + 1\big)\big(f_{,Y}+f_{,Z}\big)}
{9 (1 + \alpha)(r_f + 1)
(2r_f \epsilon_0 + 3)f}r_f\,.
\label{cL}
\ea
Since $Y$ and $Z$ are constant at the background level, both $f(Y,Z)$ and its derivatives with respect to $Y$ and $Z$ are also constant. Recalling that $r_f$ evolves as $r_f \propto a^{(3-2\epsilon_0)(1+\alpha)}$, we find that both $\bar{c}_T^2$ and $\bar{c}_L^2$ vanish in the asymptotic past, as expected. In the asymptotic future, deep in the constant-slow-roll solid regime, the propagation speeds of the transverse and longitudinal phonons approach the constant values
\begin{eqnarray}
\bar{c}_{T,\text{asy}}^2 &=& \frac{2(f_{,Y}+f_{,Z})}
{3(1+\alpha)\epsilon_0 f}\,,
\label{cTasym}\\
\bar{c}_{L,\mathrm{asy}}^2&=&
-1+\frac{2\epsilon_0}{3}+\frac{8(f_{,Y}+f_{,Z})}
{9(1+\alpha) \epsilon_0 f}\,\label{cLasym}\\
&=&-1+\frac{2\epsilon_0}{3}
+\frac{4}{3}\bar{c}_{T,\text{asy}}^2\,.
\end{eqnarray}
We can see that these expressions for the generalized Chaplygin solid satisfy the general relation \eqref{eq:cLepsilon0} 
for $\epsilon_0 \ll 1$. In particular, the expression for $\bar{c}_{L,\mathrm{asy}}^2$ shows that the solid contribution plays an essential role in ensuring the stability of perturbations in the asymptotic future constant-slow-roll phase.
For the generalized Chaplygin solid, we thus require the solid contribution to satisfy
\be
\frac{f_{,Y}+f_{,Z}}{f}\geq\frac{9}{8}(1+\alpha)\left(1-\frac{2\epsilon_0}{3}\right)\epsilon_0\simeq \frac{9}{8}(1+\alpha)\epsilon_0\,,
\ee
to avoid Laplacian instabilities in the scalar sector in the asymptotic regime. This condition is stronger than that required to avoid Laplacian instabilities of the transverse phonons, $\bar{c}_T^2\geq0$, in Eq.~\eqref{cTdef}. The combination $f_{,Y}+f_{,Z}$, which controls the importance of the solid contribution, can be expressed in terms of the more physical parameter $\bar{c}_{L,\mathrm{asy}}^2$. We can then rewrite $\bar{c}_T^2$ and $\bar{c}_L^2$ as
\ba
\bar{c}_T^2 &=& 
\frac{(3+3\bar{c}_{L,\mathrm{asy}}^2
-2\epsilon_0)\epsilon_0}{2(2\epsilon_0 r_f+3)}
r_f\,,\label{cTf}\\
\bar{c}_L^2 &=& \frac{6\epsilon_0
(r_f+1)\bar{c}_{L,\mathrm{asy}}^2
+\alpha (2\epsilon_0-3)^2}
{3 (2\epsilon_0 r_f+ 3)(r_f + 1)}r_f\,.
\label{cLf}
\ea
These expressions show that both $\bar{c}_T^2$ and
$\bar{c}_L^2$ increase from very small positive values
at early times to their asymptotic values, $c_{T,\text{asy}}^2=(3+3\bar{c}_{L,\mathrm{asy}}^2-2\epsilon_0)/4$ and $\bar{c}_{L,\mathrm{asy}}^2$, respectively. 
The asymptotic propagation speed of the transverse 
phonons takes the universal value
$\bar{c}_{T,\text{asy}}^2\simeq3/4$ for small values of
$\bar{c}_{L,\mathrm{asy}}^2$ and $\epsilon_0$, as favored by observationally viable models. 
We can see this asymptotic behaviour in Fig.~\ref{fig:cLb}. 
The fact that $\bar{c}_{T,\text{asy}}^2$ can be parametrically larger than $\bar{c}_{L,\mathrm{asy}}^2$ explains the generation of observable effects, for example in the slip parameter, while maintaining a structure formation process very similar to that of $\Lambda$CDM. At early times, the explicit evolution of the propagation speeds can be obtained by taking the limit 
$r_f \ll 1$:
\ba
\bar{c}_{T,\text{early}}^2 &\simeq& 
\frac{3+3\bar{c}_{L,\mathrm{asy}}^2
-2\epsilon_0}{6}\epsilon_0
r_f\,,\\
\bar{c}_{L,\text{early}}^2 &\simeq& \frac{6\epsilon_0
\bar{c}_{L,\mathrm{asy}}^2
+\alpha (2\epsilon_0-3)^2}
{9}r_f\,.
\label{cLf2}
\ea
Thus, at early times, both propagation speeds grow 
approximately in proportion to $r_f$.

%%%%%%%%%%%%%%%%%%%%%%%%%%%%
%%%%%%%%%%%%%%%%%%%%%%%%%%%%

Since $\epsilon_0$ needs to be a small parameter to trigger cosmic acceleration, we can expand Eq.~(\ref{cL}) 
for small $\epsilon_0$ to obtain
\be
\bar{c}_L^2(\epsilon_0 \to 0)\simeq r_f\left[\frac{\alpha}{1+r_f}+\frac{16}{27}\frac{f_{,Y}+f_{,Z}}{(1+\alpha) f}\right]\,.
\label{eq:cLsqepsilonto0}
\ee
We note that some terms proportional to $\epsilon_0$ in 
this expression are multiplied by $r_f$. Therefore, the expansion of $\bar{c}_L^2$ in small $\epsilon_0$ breaks down deep in the accelerated solid phase, where $r_f\gg1$. However, it remains valid up to the present epoch, where we expect $r_f\simeq \mathcal{O}(1)$. At early times, when $r_f\ll 1$, we have $\bar{c}_L^2\simeq r_f\left[\alpha+16(f_{,Y}+f_{,Z})
/\{27(1+\alpha)f \} \right]\ll1$ because of the smallness of $r_f$.
The stability conditions obtained above ensure stability in the asymptotic regimes. To ensure stability throughout the entire evolution, including the transition region, we need to require that Eq.~\eqref{eq:cLsqepsilonto0} remains positive for all $r_f>0$, since $r_f$ varies during cosmic evolution from $r_f \ll 1$ to $r_f \gg 1$. A sufficient condition is $\alpha>0$. Owing to the solid contribution, we can also accommodate negative values of $\alpha$ without encountering Laplacian instabilities up to the present time, provided that
\be
\frac{f_{,Y}+f_{,Z}}{f}\geq -\frac{27}{16}
\frac{\alpha (1+\alpha)}{1+r_{f,0}}\,,
\ee
where $r_{f,0}$ is the present value of $r_f$. Recalling the definition of $r_f$ in \eqref{eq:defrf}, we can estimate $r_{f,0}=(\Omega_\Lambda/\Omega_{\text{DM}})^{1+\alpha}$. For positive $\alpha$, the condition is of course satisfied upon using the stability of the transverse phonons. For negative $\alpha$, since $\alpha>-1$ is required to ensure the transition to the solid phase, we have $1\leq r_{f,0}\leq\Omega_\Lambda/\Omega_{\text{DM}}\simeq 3$. 
Stability is therefore guaranteed by imposing
\be
\frac{f_{,Y}+f_{,Z}}{f}\geq 
-\frac{27}{32}\alpha (1+\alpha).
\ee
This possibility arises only in the presence of solid contributions.

Ensuring stability is a theoretical viability condition. 
However, observational viability requires a second condition: 
$\bar{c}_L^2$ should remain sufficiently small so as not to generate strong acoustic oscillations that would otherwise be incompatible with observations. If acoustic oscillations are produced in the solid dark sector, they should appear only on relatively large scales in the future and remain under control up to the present epoch. We note that imposing $\bar{c}_{L,\text{asy}}^2\ll1$ is not by itself sufficient, because $\bar{c}_L^2$ can transiently become larger than its asymptotic value during the evolution. This behavior is illustrated in Fig.~\ref{fig:cLb}, where we show the evolution of $\bar{c}_{L}^2$ and $\bar{c}_{T}^2$.

\subsection{Solid growth and gravitational potentials: 
Numerical results}

We next study how solid-like perturbations affect the scalar sector. As a first step, we isolate the dynamics of the longitudinal mode $\Pi_L$ by neglecting radiation and baryon perturbations. This simplified setup is not intended to replace a full Boltzmann treatment, but rather to make transparent how the intrinsic sound speeds of the dark sector control the evolution of $\Pi_L$ and, consequently, the gravitational potentials. 
We numerically integrate Eq.~(\ref{pieq}) from the initial redshift $z_{\rm ini}=1000$ to today, $z=0$, neglecting the contribution from the right-hand side.
We impose the initial conditions
$\Pi_L(z_{\rm ini})=1$ and
$\dot{\Pi}_L(z_{\rm ini})=H_{\text{ini}}$, 
where $H_{\text{ini}}$ is the initial Hubble expansion rate.
Note that this choice of $\dot{\Pi}_L$ corresponds to the growing-mode solution $\Pi_L\propto a$.

\begin{figure}[t!]
\centering
\includegraphics[width=0.49\textwidth]{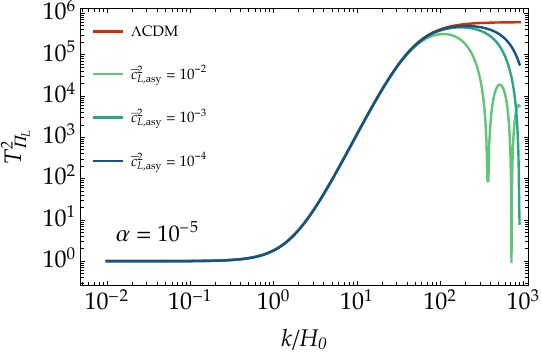}
\includegraphics[width=0.49\textwidth]{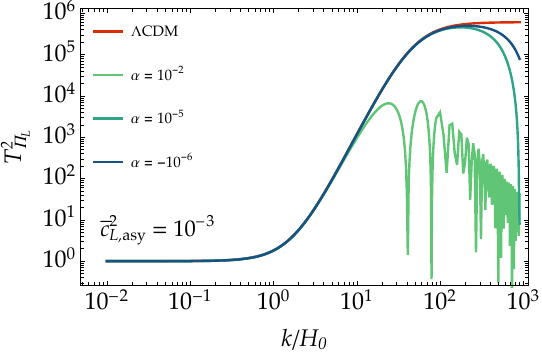}
\caption{
Transfer function $T^2_{\Pi_L}$ of the longitudinal perturbation for different values of $\bar{c}_{L,{\rm asy}}^2$ and a fixed value of $\alpha$ (left panel), and for different values of $\alpha$ and a fixed value of $\bar{c}_{L,{\rm asy}}^2$ (right panel). In both cases, we fix $\epsilon_0=0.01$ and normalize the Fourier wavenumber by the present value of the Hubble constant. The transfer functions are obtained by integrating Eq.~(\ref{pieq}), neglecting radiation and baryon perturbations, from $z_{\rm ini}=1000$ with the initial conditions 
$\Pi_L(z_{\rm ini})=1$ and $\dot{\Pi}_L(z_{\rm ini})
=H(z_{\rm ini})$. We also show the $\Lambda$CDM transfer function (red curve) as a reference model. These transfer functions clearly illustrate how acoustic oscillations can be generated on sub-Hubble scales, and how they can be pushed to arbitrarily small scales by an appropriate choice of parameters. In the right panel, we see that negative values of $\alpha$ are allowed without generating Laplacian instabilities, which would otherwise enhance the transfer function on small scales and are inevitable in the Chaplygin gas model \cite{Sandvik:2002jz}.
}
\label{figtrans}
\end{figure}

Figure~\ref{figtrans} shows the transfer function 
\be
T^2_{\Pi_L} \equiv \frac{|\Pi_L(z=0)|^2}
{|\Pi_L(z=z_{\rm ini})|^2}\,,
\ee
as a function of $k/H_0$. For sufficiently small wavenumbers,
$T^2_{\Pi_L}$ remains close to unity, and the three cases are almost indistinguishable from the flat $\Lambda$CDM reference model. For larger $k$, the curves depart from the $\Lambda$CDM behavior and begin to
oscillate. This behavior is caused by the Laplacian term
$c_L^2 k^2/a^2$ in Eq.~(\ref{pieq}), 
which provides a pressure force against the gravitational clustering of the longitudinal mode.

In the left panel of Fig.~\ref{figtrans}, we plot $T_{\Pi_L}^2$ as a function of $k/H_0$ for $\alpha=10^{-5}$ and $\epsilon_0=0.01$, considering the three representative values $\bar{c}_{L,\mathrm{asy}}^2=10^{-2}$, $10^{-3}$, and $10^{-4}$. 
Since the pressure term scales as $\bar{c}_L^2 k^2/a^2$, a larger value of $\bar{c}_L^2$ shifts the onset of oscillations to smaller wavenumbers. This explains the ordering seen in Fig.~\ref{figtrans}. On very large scales, the Laplacian term is ineffective and $T_{\Pi_L}^2\simeq1$, whereas on smaller scales the finite longitudinal rigidity produces oscillatory features and suppresses the transfer function relative to the monotonic $\Lambda$CDM-like evolution.
For late-time structure formation, scales with roughly $k<300H_0$ remain in the linear regime, while for $k>300H_0$ nonlinear corrections to matter clustering can no longer be ignored. The transfer function shown in Fig.~\ref{figtrans} should therefore be regarded as a schematic diagnostic of how the solid dark sector modifies the propagation of the longitudinal mode. In this illustrative plot, we do not include nonlinear effects associated with small-scale structure formation.

In the right panel of Fig.~\ref{figtrans}, we show the transfer function as a function of $k/H_0$ for $\bar{c}_{L,\mathrm{asy}}^2=10^{-3}$ and $\epsilon_0=0.01$, with $\alpha=10^{-2}$, $10^{-5}$, and $-10^{-6}$.
As $|\alpha|$ increases, the oscillations
of $T^2_{\Pi_L}$ start to occur at smaller values of $k$.
As $\alpha$ approaches 0, the transfer function approaches
that of $\Lambda$CDM.
As seen in the case $\alpha=-10^{-6}$, negative values
of $\alpha$ do not lead to Laplacian instabilities as long as
$|\alpha|$ is much smaller than 1. This property is
different from that of the perfect-fluid Chaplygin gas model.

%%%%%%%%%%%%%%%%%%%%%%%%%%%%
%%%%%%%%%%%%%%%%%%%%%%%%%%%%

%%%%%%%%%%%%%%%%%%%%%%%%%%%%
\begin{figure}[t!]
\centering
\includegraphics[width=0.49\textwidth]{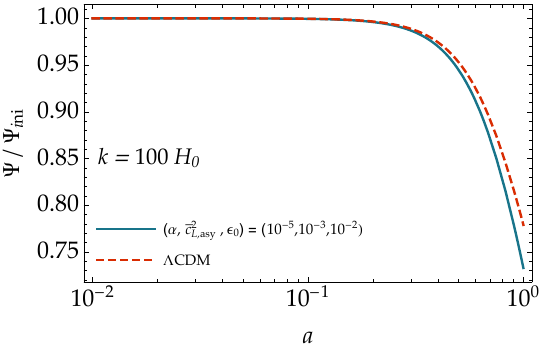}
\includegraphics[width=0.49\textwidth]{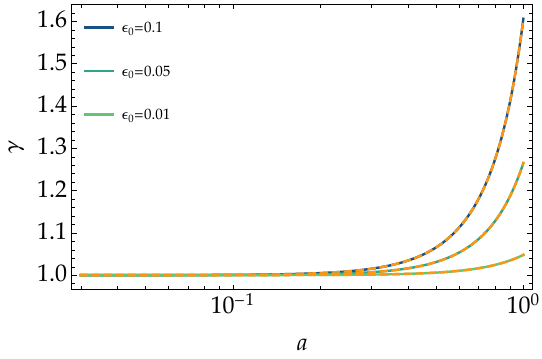}
\caption{The left panel shows the evolution of $\Psi$, normalized by its initial value
$\Psi_{\mathrm{ini}}$, for $k=100H_0$, $\alpha=10^{-5}$,
$\bar{c}_{L,\text{asy}}^2=10^{-3}$, and $\epsilon_0=0.01$.
We also plot the corresponding normalized gravitational potential in the $\Lambda$CDM limit. The gravitational potential is suppressed relative to its $\Lambda$CDM counterpart. 
In the right panel, we plot the slip parameter
$\gamma$ for $\alpha=10^{-5}$ and $\bar{c}_{L,\text{asy}}^2=10^{-3}$,
with several values of $\epsilon_0$. 
We set $k=100H_0$, although the slip is scale independent 
on sub-Hubble scales. The dashed lines represent the
analytical estimate $\gamma=(1-4\bar{c}_T^2)^{-1}$, which agrees very well with the numerical results. As explained in the main text, the solid phase can lead to observable signatures through the slip parameter.}
\label{figgra}
\end{figure}
%%%%%%%%%%%%%%%%%%%%%%%%%%%%

The evolution of the gravitational potentials is also affected by the modified growth of $\Pi_L$.
In the left panel of Fig.~\ref{figgra}, we show an example of the evolution of $\Psi$, normalized to its initial value $\Psi_{\mathrm{ini}}$, for $k=100H_0$ and fixed values of $\alpha$, $\bar{c}_{L,\text{asy}}^2$, and $\epsilon_0$. For comparison, we also plot the corresponding normalized quantity in the $\Lambda$CDM limit.
The growth of $\Pi_L$ is suppressed by the combined effect of the Laplacian term proportional to $\bar{c}_L^2 k^2/a^2$ and the mass-like contribution involving $\bar{c}_T^2$ in Eq.~(\ref{pieqsub}). This weakens the source term for the gravitational potential and suppresses $\Psi$ relative to the $\Lambda$CDM reference curve.

In the right panel of Fig.~\ref{figgra}, we plot the gravitational slip parameter for fixed values of
$\alpha$, $\bar{c}_{L,\text{asy}}^2$, and $k=100H_0$,
using three different values of $\epsilon_0$.
We observe that $\gamma$
becomes larger than unity at late times. 
This originates from the nonzero transverse 
rigidity of the dark sector.
In the sub-Hubble regime, the slip parameter can be estimated as
$\gamma=1/(1-4\bar{c}_T^2)$, which is in excellent agreement with the numerical results shown in the right panel of Fig.~\ref{figgra}.
This confirms that the deviation $\gamma-1$ is governed by the
vector-mode sound speed. We also note that the scale independence of $\gamma$ for modes with $k \gg H_0$ 
is confirmed numerically.

When the parameters $\alpha$, $\epsilon_0$, and
$\bar c_{L,{\rm asy}}^2$ are close to zero, 
both $\bar c_L^2$ and
$\bar c_T^2$ become small. 
In this limit, the longitudinal perturbation,
the gravitational potential, and the slip parameter 
evolve close to their $\Lambda$CDM counterparts. The characteristic suppression of $\Psi$ with respect to the $\Lambda$CDM case $\Psi_{\Lambda{\rm CDM}}$ 
and the enhancement of $\gamma$ found above can in principle be tested with
large-scale-structure growth-rate measurements and weak-lensing
observations. A quantitative comparison with these data, however,
requires solving the full set of perturbation equations including
radiation and baryons in a Boltzmann code. We leave such a complete
analysis for future work.

\subsection{Gravitational wave distance}
\label{secGWD}

The second-order action for tensor perturbations is given by Eq.~(\ref{St}).
Without loss of generality, we choose the wave 
vector to be along the $z=x^3$ direction and take the nonvanishing components of $h_{ij}$ as
$h_{11}=h_1(t,z)$, $h_{22}=-h_1(t,z)$, and
$h_{12}=h_{21}=h_2(t,z)$. 
Then, after Fourier transforming with respect
to $z$, the action (\ref{St}) yields the following 
equations of motion for the two tensor polarizations $h_i$ ($i=1,2$):
\be
\ddot h_i+3H\dot h_i+
\left(\frac{k^2}{a^2}+M_{\rm GW}^2
\right)h_i=0\,,
\qquad
M_{\rm GW}^2\equiv \frac{64\pi G}{9}
\left({\cal K}_{,Y}+{\cal K}_{,Z}\right)\,.
\label{hieqGW}
\ee
In conformal time, $\tau=\int a^{-1}{\rm d}t$, 
we can remove the Hubble friction 
by introducing the rescaled field
\be
\hat h_i\equiv a h_i\,.
\label{hhatdef}
\ee
Then Eq.~(\ref{hieqGW}) becomes
\be
\hat h_i''+\omega^2(\tau)\hat h_i=0\,,
\qquad
\omega^2(\tau)=k^2
+a^2 M_{\rm GW}^2(\tau)-\frac{a''}{a}\,,
\label{hieqGWconf}
\ee
where a prime denotes a derivative with respect to $\tau$,
${\cal H}=aH$, and $a''/a={\cal H}'+{\cal H}^2$. 
The coefficient of $k^2$ is the same as in $\Lambda$CDM, 
so tensor modes propagate at the speed of light. 
The new effect is the time-dependent effective mass-squared term $M_{\rm GW}^2$, 
which modifies the dispersion relation of GWs once the unified dark sector enters 
the solid phase. This effect is analogous to the effective mass acquired by photons propagating through a medium. For very high-frequency modes satisfying 
$k^2 \gg \max\{a^2M_{\rm GW}^2(\tau),a''/a\}$, this effect is negligible and GWs propagate 
as in $\Lambda$CDM. If the hierarchy $a^2M_{\rm GW}^2(\tau) \gg a''/a$ holds, 
there exists a range of sub-Hubble modes for which GWs can probe 
the effective mass generated in the solid phase.

The effective mass square of GWs can be expressed in terms of background quantities as
\be
M_{\rm GW}^2
=4\bar{c}_T^2 \epsilon_b H^2\,,
\label{MTmodel0}
\ee
which shows again that this effect is controlled by the propagation speed 
of the transverse phonons, $\bar{c}_T^2$. Using Eq.~\eqref{cTf}, we can 
further write the effective mass squared as
\be
M_{\rm GW}^2
=\frac{2(3+3\bar{c}_{L,\mathrm{asy}}^2
-2\epsilon_0)\epsilon_0}{2\epsilon_0 r_f+3}
r_f\,\epsilon_b\,H^2\,.
\label{MTmodel}
\ee
For $\epsilon_0=0$, we have $M_{\rm GW}^2=0$, and the tensor propagation equation reduces to that in $\Lambda$CDM. In the asymptotic future, where 
$\epsilon \simeq \epsilon_0 \simeq \epsilon_b$, we obtain $M_{\rm GW}^2\simeq 3\epsilon_0 H^2$ for $\bar c_{L,{\rm asy}}^2,\,\epsilon_0\ll1$, so the corresponding horizon scale deviates from that in $\Lambda$CDM only by $\mathcal{O}(\epsilon_0)$. 
This estimate is also approximately valid at the present epoch, where $r_f\lesssim\mathcal{O}(1)$ with $\epsilon_0,\bar{c}_{L,{\rm asy}}^2\ll1$. This means that, on cosmologically relevant scales, the effect of the effective mass generated by the solidity of the dark sector on the GW transfer function is negligible.

For modes satisfying the WKB condition 
$|\omega'/\omega^2|\ll1$, the
rescaled perturbation evolves adiabatically. Substituting
\be
\hat h_i(\tau,k)=A(\tau)
\exp\left[-i\int^\tau \omega(\tilde\tau)\,\rd \tilde\tau\right]
\ee
into Eq.~(\ref{hieqGWconf}) gives
$A''-2i\omega A'-i\omega' A=0$.
Since $A''$ is of higher WKB order than $\omega A'$ and $\omega'A$, we
neglect it at leading adiabatic order and obtain
$A'/A\simeq-\omega'/(2\omega)$. Thus $A\propto\omega^{-1/2}$, and the physical tensor perturbation is
\be
h_i(\tau,k)\propto
\frac{1}{a\sqrt{\omega(\tau)}}
\exp\left[-i\int^\tau \omega(\tilde\tau)\,\rd\tilde\tau\right]\,.
\label{WKBGW}
\ee
The factor $a^{-1}$ gives the standard cosmological dilution, while $\omega^{-1/2}$ is the additional propagation effect induced by the tensor
mass. In the sub-Hubble WKB regime used below,
$k^2+a^2 M_{\rm GW}^2\gg |a''/a|$, so that
\be
\omega^2(\tau) \simeq k^2+a^2M_{\rm GW}^2(\tau)\,.
\label{omegaGW}
\ee

We now define an effective GW distance by isolating this propagation
effect. We compare the observed amplitude in the present theory with the
$\Lambda$CDM amplitude for the same source, assuming that the intrinsic GW amplitude
generated at the source is unchanged from GR. With the normalization $a(z=0)=1$, Eq.~(\ref{WKBGW}) gives, for the absolute value of the amplitude
after removing the oscillating phase,
\be
h_i(0)
=
h_i(z)\,
a(z)\left[\frac{\omega(z)}{\omega(0)}\right]^{1/2}\,.
\ee
For the corresponding GW in $\Lambda$CDM, $\omega=k$, 
and hence 
\be
h_i^{\Lambda {\rm CDM}}(0)
=h_i^{\Lambda {\rm CDM}}(z)\,a(z)\,.
\ee
The source assumption gives $h_i(z)
=h_i^{\Lambda {\rm CDM}}(z)$, 
so that the standard factor $a(z)$ 
cancels in the ratio of the observed amplitudes:
\be
\frac{h_i(0)}{h_i^{\Lambda {\rm CDM}}(0)}
=
\left[\frac{\omega(z)}{\omega(0)}\right]^{1/2}.
\label{hMGRratio}
\ee
The luminosity distance $d_L$ is the distance 
inferred from the $\Lambda$CDM amplitude, 
$h_i^{\Lambda {\rm CDM}}(0)\propto d_L^{-1}$. 
We define the GW distance
$d_{\rm GW}$ analogously by 
$h_i(0)\propto d_{\rm GW}^{-1}$.
Taking the inverse of Eq.~(\ref{hMGRratio}) then yields
\be
\frac{d_{\rm GW}(z)}{d_L(z)}
=\left[\frac{\omega(0)}{\omega(z)}\right]^{1/2}
=\left[
\frac{k^2+M_{\rm GW}^2(0)}
{k^2+a^2(z)M_{\rm GW}^2(z)}
\right]^{1/4} .
\label{dgwdef}
\ee
This definition accounts only for the propagation 
correction due to the tensor mass, and does 
not include possible modifications of the source amplitude. 

Equation~(\ref{dgwdef}) shows that the correction 
is suppressed for sub-Hubble modes. 
Expanding it for $k^2\gg a^2M_{\rm GW}^2$ gives
$d_{\rm GW}/d_L-1=O(M_{\rm GW}^2/k^2)$. Astrophysical compact-binary standard
sirens are therefore insensitive to this effect. For stellar-mass
black-hole binaries observed by LIGO--Virgo, including systems with
component masses of order $10\,M_\odot$, the observed frequencies
$f_0\simeq10$--$10^3\,{\rm Hz}$ correspond to
$k/H_0\simeq3\times10^{19}$--$3\times10^{21}$. Supermassive black-hole
binaries are also far inside the horizon: pulsar-timing-array frequencies
give $k/H_0\simeq3\times10^9$--$3\times10^{11}$, while the millihertz band
of space interferometers corresponds to even larger values. For all these
sources, the deviation of $d_{\rm GW}$ from $d_L$ is negligible.

%%%%%%%%%%%%%%%%%%%%%%%%%%%%
\begin{figure*}[t!]
\centering
\includegraphics[width=0.45\textwidth]{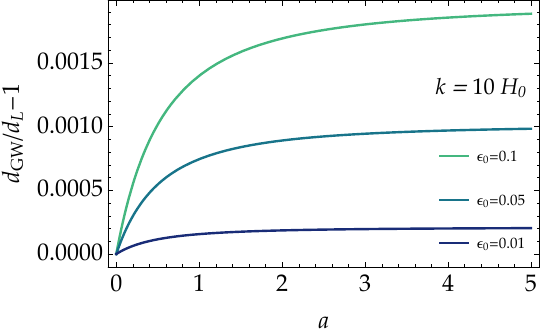}
\includegraphics[width=0.45\textwidth]{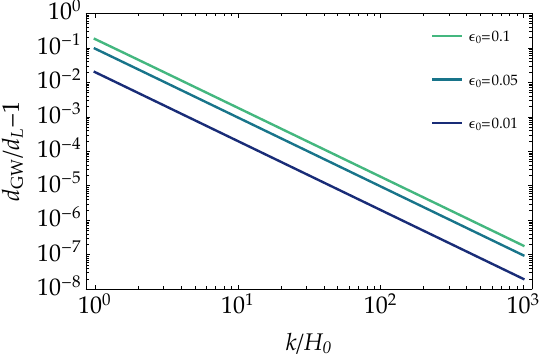}
\caption{Deviation of the GW distance $d_{\rm GW}$ from 
the standard luminosity distance $d_L$ 
for $\alpha=10^{-5}$ and $\bar{c}_{L,{\rm asy}}^2=10^{-3}$.
Left panel: redshift dependence of $d_{\rm GW}/d_L-1$
for the representative sub-Hubble tensor mode $k=10H_0$, shown for three different values of $\epsilon_0$. 
Right panel: $d_{\rm GW}/d_L-1$ at
$z=3$ as a function of $k/H_0$ in the range $1\le k/H_0\le 10^3$. 
In the right panel, the results near $k=H_0$ should be regarded only as a qualitative indication, since the WKB approximation becomes unreliable close to the horizon scale.
}
\label{figGWdistance}
\end{figure*}
%%%%%%%%%%%%%%%%%%%%%%%%%%%%

The left panel of Fig.~\ref{figGWdistance} shows that, even for the relatively large-scale sub-Hubble mode $k/H_0=10$, the deviation from the standard luminosity distance remains small.  At $z=3$, $d_{\rm GW}/d_L-1$ is below $10^{-4}$ for
$\epsilon_0 \lesssim 0.01$, and remains smaller 
than $10^{-2}$ even for optimistically large values of $\epsilon_0$ of order $0.1$.
The right panel shows the scale dependence at the same redshift. The deviation increases as $k/H_0$ decreases and can approach the percent level near $k/H_0=1$. This is precisely the regime in which the mode is close to the present horizon, so the WKB approximation is no longer quantitatively reliable. The near-horizon part of the right panel should therefore be interpreted only as a qualitative indication of the possible size of the tensor-mass propagation effect.

Modes with $k/H_0={\cal O}(1)$--${\cal O}(10)$ are not produced by local compact
binary standard sirens. They are instead naturally associated with
primordial tensor perturbations probed by CMB polarization. A tensor
multipole $\ell$ roughly corresponds to
$k/H_0\simeq \ell/(H_0\chi_*)\simeq \ell/3$, where $\chi_*$ is the
comoving distance to the last-scattering surface. Low-multipole primordial
tensor modes can therefore lie in the range where the propagation effect
is largest. They are not standard sirens, however, so the distance
interpretation of Eq.~(\ref{dgwdef}) should not be applied directly to
CMB tensor modes. A quantitative treatment of such near-horizon
primordial modes requires solving the full tensor equation rather than
using the WKB estimate.

%%%%%%%%%%%%%%%%%%%%%%%%%%%%
\section{Conclusions}
\label{conclusionsec}
%%%%%%%%%%%%%%%%%%%%%%%%%%%%

In this paper, we presented a unified dark-sector scenario in which DM
and DE arise as two regimes of a single relativistic solid. The model is
constructed to behave as pressureless matter in the early Universe and to
drive cosmic acceleration at late times. Although its background evolution
resembles that of generalized Chaplygin-type unified models, the solid
carries shear degrees of freedom, and hence its perturbation dynamics
cannot be reduced to those of a perfect fluid. We derived the background
equations, the linear stability conditions in the tensor, vector, and
scalar sectors, the gauge-ready scalar perturbation equations including
ordinary matter perturbations, and illustrative observational signatures of
this solidly unified dark sector.

In Sec.~\ref{backsec}, we formulated the background dynamics 
of the relativistic solid in the presence of ordinary matter. For the FLRW
configuration $\bar\phi^a=\lambda x^a$, only the compression invariant $b$ evolves with the expansion, while the shape invariants $Y$ and $Z$ remain
constant on the background. We then introduced a generalized
Chaplygin-solid realization in which the dark-sector density scales as
CDM at early times and approaches a slowly varying DE-like regime in
the future. The quantity $r_f$ controls the transition between these two
regimes. For $r_f\ll1$, the dark component has $w_{\rm D}\simeq0$, leading
to a standard matter-dominated epoch after radiation domination. When
$r_f$ becomes of order unity, $w_{\rm D}$ decreases and the Universe
enters a phase of accelerated expansion. We also clarified how the
unified case with $\alpha\ne0$ differs from the separable $\alpha=0$ case,
in which the Lagrangian can be decomposed into dustlike and
DE-like pieces.

In Sec.~\ref{sec:stability}, we derived the linear stability conditions by
computing the quadratic action in the regime dominated by the dark solid,
without including the matter Lagrangian ${\cal L}_{\rm M}$. This is
sufficient for the stability analysis, because ordinary matter is
minimally coupled and mixes with the dark sector only through gravity,
which decouples on small scales. In the tensor sector, we found that GWs
propagate at the speed of light but acquire a mass term sourced by the
shear response of the solid. In the vector sector, after integrating out
nondynamical variables, the transverse phonon modes were shown to be the
propagating degrees of freedom. Their no-ghost condition requires
$b{\cal K}_{,b}>0$, while Laplacian stability requires a positive vector sound speed squared. 
In the scalar sector of the dark solid, the longitudinal phonon is the single dynamical scalar degree of freedom. 
The absence of
Laplacian instabilities requires a positive 
longitudinal sound speed squared,
which contains both the perfect-fluid contribution and the additional
solid contribution controlled by the shear response.

In Sec.~\ref{perturbsec}, we reinstated the matter sector and derived the
linear scalar perturbation equations in a gauge-ready form, following the
four-subsection structure of that section. We first introduced
gauge-invariant variables for the gravitational potentials, the phonons,
and the matter perturbations. We then derived the gravitational equations
with contributions from both the solid dark sector and ordinary matter,
showing that the solid anisotropic stress generates a gravitational slip,
$\Psi\ne\Phi$, controlled by the same combination
${\cal K}_{,Y}+{\cal K}_{,Z}$ that determines the vector sound speed and
the tensor mass. We next combined the longitudinal phonon equation with
the gravitational and matter equations to study solid growth. In the deep
matter era, after neglecting subdominant baryon and radiation sources, the
sound speeds are suppressed and the growing solution behaves as
$\Pi_L\propto a$, giving matterlike clustering. At low redshifts, the
longitudinal sound speed provides pressure support, while the vector sound
speed weakens the effective gravitational coupling and produces the slip.
Finally, in a simplified low-redshift treatment, baryon perturbations obey
the standard continuity and Euler equations and act as nonrelativistic
tracers of the potential sourced mainly by the solid dark sector.

In Sec.~\ref{signaturemodel}, we illustrated the observational signatures
of the generalized Chaplygin solid, following the three subsections of
that section. We first identified the theoretically viable region by
expressing the vector and longitudinal sound speeds in terms of $r_f$ and
the asymptotic sound speed. Both speeds vanish in the asymptotic past and
approach constant values in the solid phase; stability requires a
sufficient solid response, while observational viability requires the
longitudinal sound speed to remain small enough to avoid strong acoustic
oscillations on structure-formation scales. We then studied the numerical
evolution of the longitudinal mode and gravitational potentials in this
regime. The transfer function of $\Pi_L$ remains close to the
$\Lambda$CDM-like reference behavior on large scales, whereas finite
longitudinal rigidity induces oscillations and suppression on smaller
scales. The same solid response suppresses the gravitational potential and
produces a nonzero, nearly scale-independent slip at low redshifts, a
signature absent in perfect-fluid Chaplygin-gas models. Finally, we
studied GW propagation. The shear response gives tensor modes a
time-dependent mass, modifying the GW amplitude and leading to a GW
distance different from the standard luminosity distance. This effect is
negligible in the asymptotic past and becomes relevant only at low redshifts, when the DE-like solid phase is important.

The next step is to confront this framework with cosmological data in a
fully quantitative manner. This requires implementing the gauge-ready
perturbation equations in a Boltzmann code, including baryons, radiation,
photon-baryon coupling, and the full hierarchy of relativistic species,
and performing a Markov-chain-Monte-Carlo analysis of the model parameters. 
It will be particularly important to combine current CMB, supernova, BAO,
weak-lensing, and redshift-space-distortion data, including recent DESI
measurements, and to extend the analysis to future surveys such as Euclid
and the Vera Rubin Observatory. Such an analysis will determine whether
the solid unified dark sector developed here can provide an observationally
viable alternative to the standard description in terms of separate DM and
DE components.

%%%%%%%%%%%%%%%%%%%%%%%%%%%
\section*{Acknowledgements}
%%%%%%%%%%%%%%%%%%%%%%%%%%%%

J.B.J and F.A.T.P. thank the hospitality of the Department of Physics of Waseda University during a visit where this work was initiated. S.T. thanks the members of the University of Salamanca for their warm hospitality during the period in which part of this work was carried out. This work has been supported by the grants PID2024-158938NB-I00 funded by MICIU/AEI/10.13039/ 501100011033 and by “ERDF A way of making Europe” and the Project SA097P24 funded by Junta de Castilla y Le\'on. M.P.G. acknowledges support from Fondo Social Europeo Plus, Programa Operativo de Castilla y León and Junta de Castilla y León, through the Consejería de Educación. 
S.T. acknowledges support from JSPS KAKENHI Grant Nos.~26K07090 and 26H00847, and from the Waseda University Special Research 
Projects (No.~2026C-486). 

\appendix

\section{Second-order action for vector and scalar perturbations}

In this Appendix, we present the quadratic-order action obtained by expanding Eq.~(\ref{action}) up to second order 
in vector and scalar perturbations and study the linear stability conditions in the presence of the matter Lagrangian ${\cal L}_{\rm M}$.
For this purpose, we describe the matter action 
${\cal S}_{\rm M}=\int {\rm d}^4 x \sqrt{-g}\,{\cal L}_{\rm M}$ 
for a perfect fluid in the Schutz-Sorkin form \cite{Schutz:1977df,Brown:1992kc,DeFelice:2009bx}
\be
{\cal S}_{\rm M}=-\int {\rm d}^4 x \left[ \sqrt{-g}\,\rho_{\rm M}(n)
+J^{\mu} \left( \partial_{\mu} \ell +{\cal A}_1 \partial_{\mu} 
{\cal B}_1+{\cal A}_2 \partial_{\mu} 
{\cal B}_2\right) \right]\,,
\label{calSM}
\ee
where $\rho_{\rm M}$ is the fluid energy density, which depends on the number density $n$. We focus on a single fluid here, but the extension to multiple perfect fluids is straightforward.
The current vector field $J^{\mu}$ in Eq.~(\ref{calSM}) is related to $n$ through 
$n=\sqrt{J_{\mu}J^{\mu}/g}$, while $\ell$ is a Lagrange multiplier. 
The quantities ${\cal A}_1$, ${\cal A}_2$, ${\cal B}_1$, and ${\cal B}_2$ 
account for the contributions from intrinsic vector modes. 
In the tensor sector, the final second-order action is of the form (\ref{St}) 
even in the presence of the matter action (\ref{calSM}). 
In the following, we study vector and 
scalar perturbations in turn.

\subsection{Vector perturbations}

In the vector sector, the perturbed line element (\ref{perlineN}) contains only the perturbed field $S_i$ after imposing the gauge in which the contribution $2a^2 \partial_i F_j \rd x^i \rd x^j$ vanishes.
Choosing the Fourier wave vector to be along the $z$ direction, we can take the nonvanishing components of $S_i$ and 
the transverse phonon perturbation $\pi_T^i$ to be
\be
S_i=\left[ S_1(t,z), S_2(t,z), 0 \right]\,,\qquad
\pi_T^i=\left[ \pi_{T1}(t,z), \pi_{T2}(t,z), 0\right]\,,
\ee
In the matter sector described by the action (\ref{calSM}), 
we take the following configurations
for ${\cal A}_1$, ${\cal A}_2$, ${\cal B}_1$, ${\cal B}_2$,
and for the spatial components of $J^{\mu}$ \cite{DeFelice:2009bx}:
\ba
& &
{\cal A}_1=\delta {\cal A}_1(t,z)\,,\qquad
{\cal A}_2=\delta {\cal A}_2(t,z)\,, \qquad
{\cal B}_1=x+\delta {\cal B}_1 (t,z)\,,\qquad
{\cal B}_2=y+\delta {\cal B}_2 (t,z)\,,\nonumber \\
& &
J^i=\frac{1}{a^2} \delta^{ik}W_k\,,\quad {\rm with} \quad
W_j=[W_1(t,z), W_2(t,z), 0]\,,
\ea
where $\delta {\cal A}_1$, $\delta {\cal A}_2$,
$\delta {\cal B}_1$, $\delta {\cal B}_2$, $W_1$,
and $W_2$ represent perturbations.
Expanding the Schutz-Sorkin action (\ref{calSM}) up to second
order in vector perturbations and integrating out the fields
$\delta {\cal A}_i$ and $W_i$, we obtain the following
reduced action \cite{DeFelice:2016yws,DeFelice:2016uil} \footnote{
This corresponds to setting $V_i=-aS_i$ in Eq.~(3.24) 
of Ref.~\cite{DeFelice:2016yws}.}
\be
({\cal S}_{\rm M}^{(2)})_v = -\int {\rm d}^4 x
\sum_{i=1}^2 \frac{a^3}{2} \left[ (\rho_{\rm M}
+P_{\rm M}) \left( S_i+a \dot{\delta {\cal B}}_i
\right)^2-\rho_{\rm M} S_i^2 \right],
\ee
We also expand the action ${\cal S}_{R{\cal K}}=\int {\rm d}^4 x \sqrt{-g}
\left[ R/(16 \pi G)-{\cal K}(b,Y,Z) \right]$ 
up to second order and perform several integrations by parts. 
This procedure leads to the quadratic action
\ba
({\cal S}_{R{\cal K}})_v^{(2)}
&=& \int {\rm d}^4 x \sum_{i=1}^2 
\biggl[ \frac{\lambda^6 {\cal K}_{,b}}{a} 
(\dot{\pi}_{Ti})^2
-\frac{2}{9}a^3 ({\cal K}_{,Y}+{\cal K}_{,Z}) (\partial \pi_{Ti})^2
+\frac{2\lambda^6 {\cal K}_{,b}}{a^2}
\dot{\pi}_{Ti} S_i \nonumber \\
& &\qquad\qquad\quad
+\frac{a^4 (\partial S_i)^2-2 \{ 8 \pi G (a^6 {\cal K}-2\lambda^6
{\cal K}_{,b})-3a^6 H^2 \}S_i^2}{32\pi G a^3} \biggr].
\ea
The total action for vector perturbations is given by 
${\cal S}_v^{(2)}=({\cal S}_{\rm M}^{(2)})_v+({\cal S}_{R{\cal K}})_v^{(2)}$.
We move to Fourier space and decompose a generic perturbation
${\cal X}$ as in Eq.~(\ref{calX}), omitting the subscript $k$ 
for perturbations in Fourier space. 
We then vary
${\cal S}_v^{(2)}$ with respect to the nondynamical perturbations
$S_i$ and $\delta{\cal B}_i$ and integrate them out. Using the
background Eq.~(\ref{back1}) and the relation $\lambda=b^{1/6}a$,
we obtain the reduced action
\be
{\cal S}_v^{(2)}=\int \frac{{\rm d}t {\rm d}^3 k}{(2\pi)^3}
\sum_{i=1}^2 a^5\frac{b {\cal K}_{,b}}{1+4\epsilon_b a^2H^2/k^2}
\left[(\dot{\pi}_{Ti})^2-c_T^2\frac{k^2}{a^2} \pi_{Ti}^2\right]\,,
\ee
where $\epsilon_b$ and $c_T^2$ are given in Eq.~(\ref{cTdef}). 
This second-order action has the same form as the corresponding action
derived without the matter action ${\cal S}_{\rm M}$.
Therefore, even in the presence of matter, including baryons and radiation, 
the conditions for the absence of ghosts and Laplacian instabilities 
are still given by Eqs.~(\ref{NGvec}) and (\ref{Lapvec}).

\subsection{Scalar perturbations}

For scalar perturbations, we do not fix the gauge 
conditions from the 
outset, but instead derive the second-order action 
in a gauge-ready form by using the perturbed 
line element (\ref{perline}).
In the Schutz-Sorkin action~(\ref{calSM}), the current $J^{\mu}$ contains 
two scalar perturbations, $\delta J$ and $\delta j$, as 
$J^0 = {\cal N}_0 + \delta J$ and 
$J^i = \delta^{ik} \partial_k \delta j / a^2$, where 
${\cal N}_0 = n_0 a^3$ and $n_0$ is the background 
fluid number density. The fluid four-velocity is given by 
$u^{\mu} = J^{\mu}/(n\sqrt{-g}) = \partial^{\mu} \ell / \rho_{{\rm M},n}$, 
where $\rho_{{\rm M},n} = \partial \rho_{\rm M} / \partial n$. 
The spatial components of $u^{\mu}$ can be expressed as 
$u^{i} = -\partial^i v$, where $v$ is the velocity potential. 
Then, the Lagrange multiplier $\ell$ takes the form 
$\ell=-\int^t \rho_{{\rm M},n} (\tilde{t}){\rm d} \tilde{t}-\rho_{{\rm M},n}v$. 
We also define the matter density perturbation as 
\be
\delta \rho_{\rm M}=\frac{\rho_{{\rm M},n}}{a^3} 
\left[ \delta J -{\cal N}_0 \left( 3 \zeta+\partial^2 E 
\right) \right]\,,
\ee
where $\partial^2 \equiv \delta^{ij} \partial_i \partial_j$. 
With this definition of $\delta \rho_{\rm M}$, the perturbation of 
the fluid number density, $\delta n$, coincides with 
$\delta \rho_{\rm M} / \rho_{{\rm M},n}$ at linear order. 
Expanding this matter action up to second order in 
scalar perturbations and integrating out the field $\delta j$, 
the resulting quadratic-order action takes the form \cite{Kase:2018aps}
\ba
\hspace{-0.7cm}
({\cal S}_{\rm M}^{(2)})_s 
&=& \int {\rm d}^4 x\,a^3 \biggl[  
\left( \dot{v}-3H c_{\rm M}^2 v-A\right) \delta \rho_{\rm M}
-\frac{c_{\rm M}^2}{2(\rho_{\rm M}+P_{\rm M})} \delta \rho_{\rm M}^2 
-\frac{\rho_{\rm M}+P_{\rm M}}{2a^2} 
\left\{ (\partial v)^2+2a \partial v \partial \chi \right\}
-\frac{\rho_{\rm M}}{2} (\partial \chi)^2 \nonumber \\
\hspace{-0.7cm}
&&\qquad \qquad\, +\frac{\rho_{\rm M}}{2}A^2 
+\frac{P_{\rm M}}{2} \left( \zeta+\partial^2 E \right) 
\left( 3\zeta-\partial^2 E \right)+\left( 3\zeta+\partial^2 E \right)
\left\{ (\rho_{\rm M}+P_{\rm M}) \left( \dot{v}-3H c_{\rm M}^2 v \right)
-\rho_{\rm M}A \right\} \biggr]\,,
\ea
where $c_{\rm M}^2$ is the squared matter sound 
speed defined in Eq.~(\ref{cM}). 
Since the background fluid pressure is expressed 
as $P_{\rm M}=n_0 \rho_{{\rm M},n}-\rho_{\rm M}$, 
we can also express $c_{\rm M}^2$ as 
$c_{\rm M}^2=n_0 \rho_{{\rm M},nn}/\rho_{{\rm M},n}$.

Expanding the action ${\cal S}_{R{\cal K}}=\int {\rm d}^4 x \sqrt{-g}
\left[ R/(16 \pi G)-{\cal K}(b,Y,Z) \right]$ up to 
second order in scalar perturbations and performing 
integration by parts, the resulting 
quadratic-order action takes the form
\be
({\cal S}_{R {\cal K}})^{(2)}=\int {\rm d}^4 x 
\left( {\cal L}_{\rm flat}+{\cal L}_{\zeta}
+{\cal L}_{E} \right)\,,
\ee
where 
\ba
{\cal L}_{\rm flat} &=& \frac{\lambda^6{\cal K}_{,b}}{a} 
\dot{\partial \pi}_L^2-\frac{27\lambda^6 (a^6{\cal K}_{,b} 
+ 2\lambda^6 {\cal K}_{,bb})+8a^{12}({\cal K}_{,Y}
+{\cal K}_{,Z})}{27 a^9} (\partial^2 \pi_L)^2
-\frac{2\lambda^6 {\cal K}_{,b}}{a^3} \left( a \partial \chi 
\dot{\partial \pi}_L+A \partial^2 \pi_L \right) \nonumber \\
& &+\frac{a^3(8 \pi G {\cal K} - 9H^2)}{16 \pi G}A^2
-\frac{a^2 H}{4\pi G}A \partial^2 \chi
-\frac{8\pi G(a^6 {\cal K} - 2\lambda^6 
{\cal K}_{,b}) - 3a^6H^2}{16 \pi G a^3} (\partial \chi)^2\,,\\
{\cal L}_{\zeta} &=& -\frac{1}{8\pi G} \bigg[ 
3a^3 \dot{\zeta}^2 - a (\partial \zeta)^2 
+\frac{3 \{ (8 \pi G {\cal K} - 3H^2 - 2 \dot{H}) a^{12} 
+ 32\pi G \lambda^6 (a^6 {\cal K}_{,b} 
+ 3\lambda^6 {\cal K}_{,bb})\}}{2a^9} \zeta^2
\biggr]+\frac{a^2}{4\pi G} \dot{\zeta}\partial^2 \chi 
\nonumber \\
& &-\frac{1}{8\pi G} \left[ 2a \partial^2 \zeta
+a^3 (24 \pi G {\cal K} \zeta-6 H \dot{\zeta}
-9 H^2 \zeta)-\frac{48 \pi G \lambda^6 {\cal K}_{,b}
\zeta}{a^3} \right] A+\frac{6\lambda^6}{a^9} 
\left( a^6 {\cal K}_{,b}+2\lambda^6 {\cal K}_{,bb} 
\right)\zeta \partial^2 \pi_{L}\,,\\
{\cal L}_{E} &=& 
-\left[ \frac{2\lambda^6{\cal K}_{,b}}{a^3}
+\frac{2\lambda^{12}{\cal K}_{,bb}}{a^9} 
+\frac{a^3\{81 H^2+54 \dot{H}-8 \pi G 
(27 {\cal K}-16{\cal K}_{,Y}-16{\cal K}_{,Z})\}}{432 \pi G}
\right] (\partial^2 E)^2 \nonumber \\
& &-a^3 \left[ {\cal K} (A+\zeta)
-\frac{16}{27} ({\cal K}_{,Y}+{\cal K}_{,Z}) \partial^2 \pi_L 
\right] \partial^2 E+\frac{2\lambda^6}{a^3} 
\left[ {\cal K}_{,b}(A-2\zeta+\partial^2 \pi_L)
-\frac{2\lambda^6 {\cal K}_{,bb}}{a^6} 
(3\zeta-\partial^2 \pi_L)\right] \partial^2 E \nonumber\\
& &+\frac{a^3}{8\pi G} \left[ 
2 \ddot{\zeta} - 2H(\dot{A}-3 \dot{\zeta}) 
-(3 H^2 + 2 \dot{H})(A - \zeta)\right] \partial^2 E\,.
\ea
The equations of motion for $A$, $\chi$, $\zeta$, $E$, 
$\pi_L$, $v$, and $\delta \rho_{\rm M}$ follow from varying 
the total second-order action 
${\cal S}_s^{(2)} = ({\cal S}_{R {\cal K}})^{(2)} + ({\cal S}_{\rm M}^{(2)})_s$ with respect to these fields.
They can be expressed in terms of the gravitational 
potentials $\Psi$ and $\Phi$, 
the longitudinal scalar-field perturbation $\Pi_L$ defined in Eq.~(\ref{Grapo}), and 
the gauge-invariant perturbations 
$\delta \rho_{\rm N}=\delta \rho_{\rm M}
+a \dot{\rho}_{\rm M} ( \chi - a \dot{E} )$ and 
$v_{{\rm N}} = v + a ( \chi - a \dot{E} )$. 
The resulting perturbation equations of motion 
in Fourier space coincide with those given in 
Eqs.~(\ref{Eq:dG00})-(\ref{Eq:dGan}) and (\ref{Eq:PiL}). 
Each matter species also obeys the continuity and Euler equations, (\ref{pereq5D}) 
and (\ref{pereq6D}).
We recall that these equations are written in a gauge-ready form.

Since the gauge transformation contains two scalar gauge degrees of freedom, 
associated with $\xi^0$ and $\xi$, we can set two of the perturbation variables 
to zero. Some examples of gauge choices are as follows:
\begin{itemize}
\item
(i) Newtonian gauge: $\chi = 0$ and $E = 0$, for which $\Psi = A$, 
$\Phi = -\zeta$, $\Pi_L = \pi_L$, $\delta \rho_{\rm N}=\delta \rho_{\rm M}$, 
and $v_{\rm N}=v$, 
\item
(ii) Flat gauge: $\zeta = 0$ and $E = 0$, for which 
$\Psi = A + (a\chi)^{\cdot}$, $\Phi = -aH \chi$, $\Pi_L = \pi_L$, 
$\delta \rho_{\rm N} = \delta \rho_{\rm M} 
+ a \dot{\rho}_{\rm M} \chi$, 
$v_{\rm N} = v + a \chi$,
\item
(iii) Synchronous gauge: $A = 0$ and $\chi = 0$, for 
which $\Psi = -(a^2 \dot{E})^{\cdot}$, $\Phi = -\zeta + a^2 H \dot{E}$, $\Pi_L = \pi_L - E$, 
$\delta \rho_{\rm N} = \delta \rho_{\rm M}-a^2 
\dot{\rho}_{\rm M} \dot{E}$, and $v_{\rm N} = v-a^2 \dot{E}$.
\end{itemize}
Let us derive conditions for the absence of ghosts and 
Laplacian instabilities in the small-scale 
limit by choosing the flat gauge ($\zeta=0=E$). 
Eliminating the nondynamical fields 
$A$, $\chi$, $v$ from the second-order action 
${\cal S}_s^{(2)}$ by using their perturbation equations 
of motion, the reduced action in Fourier space 
takes the form 
\be
{\cal S}_s^{(2)}=\int \frac{{\rm d} t{\rm d}^3 k}{(2\pi)^3}a^{3}\left( 
\dot{\vec{\mathcal{X}}}^{t}{\bm K}\dot{\vec{\mathcal{X}}}
-\vec{\mathcal{X}}^{t}{\bm G}\vec{\mathcal{X}}
-\vec{\mathcal{X}}^{t}{\bm B}\dot{\vec{\mathcal{X}}}
\right)\,,
\ee
where ${\bm K}$ and ${\bm G}$ are $2 \times 2$ symmetric matrices, ${\bm B}$ is an antisymmetric matrix, and 
$\vec{\mathcal{X}}^{t}$ is composed of the two dynamical perturbations $\Pi_L$ and $\delta \rho_{\rm N}$ as
\be
\vec{\mathcal{X}}^{t}=\left( k \Pi_L, 
\delta \rho_{\rm N}/k \right) \,.
\ee
At leading order in the expansion with large values of 
$k$, the matrix components are given by 
\ba
& &
K_{11}=a^2 b {\cal K}_{,b}+{\cal O}(k^{-2})\,,\quad 
K_{22}=\frac{a^2}{2(\rho_{\rm M}+P_{\rm M})}
+{\cal O}(k^{-2})\,,\quad 
K_{12}=K_{21}=12\pi G a^4 b {\cal K}_{,b}k^{-2}
+{\cal O}(k^{-4}),\nonumber \\
& &
G_{11}=\bar{c}_L^2 b {\cal K}_{,b} k^2+{\cal O}(k^0)\,,
\quad 
G_{22}=\frac{c_{\rm M}^2 k^2}{2(\rho_{\rm M}
+P_{\rm M})}+{\cal O}(k^0)\,,
\quad 
G_{12}=G_{21}=2\pi G a^2 b \left[ (4+3c_{\rm M}^2)
{\cal K}_{,b}+6b{\cal K}_{,bb} \right]+{\cal O}(k^{-2}),
\nonumber \\
& &
B_{12}=-B_{21}=-36(1+c_{\rm M}^2) \pi G a^7 H 
b {\cal K}_{,b}k^{-2}+{\cal O}(k^{-4})\,.
\ea

The absence of ghosts requires that $K_{11} > 0$ 
and ${\rm det}\,{\bm K} = K_{11}K_{22} - K_{12}^2 > 0$.
Since $K_{12}$ is suppressed relative to $K_{11}$ and $K_{22}$ in the large-$k$ limit, the ghost-free conditions reduce to $K_{11} > 0$ and $K_{22} > 0$, namely
\be
b {\cal K}_{,b}>0\,,\qquad 
\rho_{\rm M}+P_{\rm M}>0\,.
\label{scasta1}
\ee
The former is equivalent to the no-ghost condition for vector perturbations.

Since the matrix components in ${\bm M}$ are suppressed 
compared to those in ${\bm K}$ and ${\bm G}$ for 
large values of $k$, the squared propagation speeds of 
$c_s^2$ are known by solving 
\be
{\rm det}\,\left( c_s^2 \frac{k^2}{a^2} {\bm K}-{\bm G} 
\right)=0\,.
\ee
Taking the large-$k$ limit, we obtain the two solutions 
$c_s^2=\bar{c}_L^2$ and $c_s^2=c_{\rm M}^2$. 
To avoid Laplacian instabilities, we require that 
\be
\bar{c}_L^2>0\,,\qquad c_{\rm M}^2>0\,.
\label{scasta2}
\ee

The linear stability conditions (\ref{scasta1}) and (\ref{scasta2}) have been derived in the flat gauge, 
but they also hold for other gauges in which $\xi^0$ and $\xi$ are completely fixed.
Since the off-diagonal components of the matrices
${\bm K}$, ${\bm G}$, and ${\bm B}$ affect neither the ghost-free conditions nor 
the scalar propagation speeds in the large-$k$ limit,
the two dynamical perturbations $\Pi_L$ and $\delta \rho_{\rm N}$ propagate independently.
This implies that the right-hand side of Eq.~(\ref{pieq}) can be neglected compared to the left-hand side for perturbations deep inside the Hubble radius ($k \gg aH$).
For modes closer to the Hubble radius, however, we need to incorporate the contributions from the perturbations $\delta \rho_{\rm N}$ and $v_{\rm N}$ to accurately estimate the evolution of $\Pi_L$.
Under the ghost-free condition $b {\cal K}_{,b} > 0$, the parameter $\epsilon_b$ is positive, and hence $c_L^2 > 0$ as long as $\bar{c}_L^2 > 0$.

\bibliographystyle{mybibstyle}
\bibliography{Solidbib}

\end{document}